\documentclass[fleqn,10pt]{wlscirep}
\usepackage[utf8]{inputenc}
\usepackage[T1]{fontenc}
\usepackage{mdframed}
\usepackage{soul}
\usepackage{academicons}
\usepackage{fontawesome5}
\usepackage{amsmath}

\newcommand{\apj}   {{Astrophys. J.}} % {{ApJ}}
\newcommand{\apjl}   {{Astrophys. J. Lett.}}
\newcommand{\apjs}  {{Astrophys. J., Suppl. Ser.}}
\newcommand{\nat}    {{Nature}}
\newcommand{\solphys} {{Sol. Phys.}}

\newcommand{\aapr} {{Astron. Astrophys. Rev.}}
\newcommand{\aap}  {{Astron. Astrophys.}}

\newcommand{\mnras}  {{Mon. Not. R. Astron. Soc.}}
\newcommand{\planss}  {{Planet. Space Sci.}}
\newcommand{\sovast}  {{Sov. Astron.}}
\newcommand{\jgr}  {{J. Geophys. Res.}}
\newcommand{\pasj}  {{Publ. Astron. Soc. Jpn.}}
\newcommand{\pasp}  {{Publ. Astron. Soc. Pac.}}
\newcommand{\aj}  {{Astron. J.}}
\newcommand{\baas}  {{Bull. Am. Astron. Soc.}}

\newcommand{\orcid}[1]{\href{https://orcid.org/#1}{\textcolor[HTML]{A6CE39}{\faOrcid}}}

\newcommand{\SYS}[1]{{#1}}

\newcommand{\BC}[1]{{#1}}

\title{Discovery of long-lasting auroral-like radio emission above a sunspot}

\author[1,*]{Sijie Yu \orcid{0000-0003-2872-2614}}
\author[1]{Bin Chen \orcid{0000-0002-0660-3350}}
\author[2]{Rohit Sharma \orcid{0000-0003-0485-7098}}
\author[3]{Timothy S. Bastian \orcid{0000-0002-0713-0604}}
\author[1]{Surajit Mondal \orcid{0000-0002-2325-5298}}
\author[1]{Dale E. Gary \orcid{0000-0003-2520-8396}}
\author[1,4]{Yingjie Luo \orcid{0000-0002-5431-545X}}
\author[2]{Marina Battaglia \orcid{0000-0003-1438-9099}}
\affil[1]{New Jersey Institute of Technology, 323 M.L.K. Blvd, Newark, NJ 07102, USA}
\affil[2]{University of Applied Sciences and Arts Northwestern Switzerland, 5210 Windisch, Switzerland}
\affil[3]{National Radio Astronomy Observatory, 520 Edgemont Road, Charlottesville, VA 22903, USA}
\affil[4]{School of Physics \& Astronomy, University of Glasgow, Glasgow, G12 8QQ, UK}

\affil[*]{sijie.yu@njit.edu}

\begin{abstract}

Auroral radio emissions in planetary magnetospheres typically feature highly polarized, intense radio bursts, usually attributed to electron cyclotron maser (ECM) emission from energetic electrons in the planetary polar region \BC{that features a converging magnetic field}. Similar bursts have been observed in magnetically active low-mass stars and brown dwarfs, often prompting analogous interpretations. Here we report observations of long-lasting solar radio bursts with high brightness temperature, wide bandwidth, and high circular polarization fraction akin to these auroral/exo-auroral radio emissions, \SYS{albeit two to three orders of magnitude weaker}  \BC{than those on certain low-mass stars}. Spatially, spectrally, and temporally resolved analysis suggests that the source is located above a sunspot where a strong, converging magnetic field is present. The source morphology and frequency dispersion are consistent with ECM emission due to precipitating energetic electrons produced by recurring flares nearby. Our findings offer new insights into the origin of such intense solar radio bursts and \BC{may provide} an alternative explanation for auroral-like radio emissions on other flare stars with large starspots.

\end{abstract}
\begin{document}

\flushbottom
\maketitle

\thispagestyle{empty}

\section*{Main text}

Stellar magnetism is not only essential for exploring the fundamentals of stellar magnetic activity and internal dynamo, but also for understanding its influence on stellar ecosystems, including potentially habitable exoplanets. Radio observations of coherent emission offer unique diagnostics of physical parameters in stellar atmospheres. Specifically, radio electron cyclotron maser (ECM) emission occurs at low harmonics of the gyrofrequency $\nu_\mathrm{B}$, providing direct measurements of magnetic field strengths above the surface of stars. Radio ECM bursts, commonly observed in planetary magnetospheres, are attributed to the ECM instability driven by the precipitation of energetic electrons to the polar regions of the planets\cite{1998JGR...10320159Z}. These radio bursts are distinguished by their highly polarized, broadband, and strongly beamed nature. Similar, albeit more intense, radio emissions have been identified in low-mass stars\cite{2005ApJ...627..960B}. In particular, recent evidence of periodic broadband radio pulses with high circular polarization from M-dwarfs \cite{2005ApJ...627..960B,2007ApJ...663L..25H,2022ApJ...935...99B} favor the planetary hypothesis that these stellar radio bursts are radio ECM emission generated in a predominately dipolar field's low-density cavity above the polar regions, in synchrony with the stellar rotation \cite{2015Natur.523..568H,2019MNRAS.488..559Z}. If interpreted accurately, these findings offer a new avenue for probing magnetic fields in the coronae of stellar and sub-stellar objects, a critical parameter for understanding their internal dynamo and examining exo-space weather and habitability\cite{2019BAAS...51c.484K}.

A key challenge in comprehending these stellar radio bursts is characterizing the emission source region due to the absence of spatially resolved observations, except for rare instances from the Very Long Baseline Interferometry (VLBI) observations \cite{1998A&A...331..596B,2023arXiv230212841K}. As the magnetic field topologies of radio-ECM-emitting stars remain largely undetermined, it is debated whether these emissions originate from a global dipole magnetic field driven by auroral-like activities in the magnetosphere\cite{2009ApJ...695..310B}, or from localized magnetic field structures, such as starspots, driven by flare-like magnetic activities in the coronae\cite{2011A&A...525A..39Y,2012ApJ...746...99K}. The Sun, owing to its proximity, offers valuable context for studying radio ECM emissions analogous to those in (sub)stellar systems. Thorough analysis of such radio emissions from the Sun could inform our understanding of other stars and carry significant implications for uncovering the origins of intense radio bursts in (sub)stellar systems, particularly those with sizable starspots. However, while transient radio ECM bursts have been sporadically recorded on the Sun\cite{2019NatCo..10.2276C}, persistent radio ECM emissions akin to those documented in stellar literature have yet to be observed on the Sun.

\section*{Results}
\begin{figure}[!htb]
\centering
\includegraphics[width=0.8\linewidth]{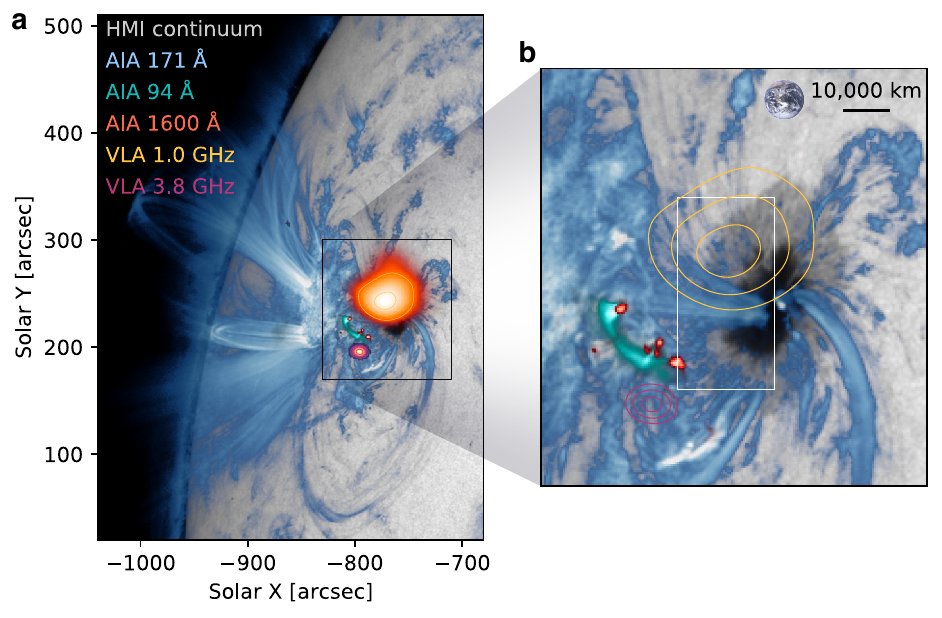}
\caption{\textbf{Auroral-like radio source seen from a sunspot associated with solar flares.} \textbf{a}, Example of a VLA snapshot image at 1.0 GHz (orange) of the radio emission from the sunspot with the image of the radio-hosting NOAA 12529 AR observed in EUV wavelengths by the Atomospheric Imaging Assembly (AIA) 171 \AA\ (blue), 94 \AA\ (cyan) and 1600 \AA\ (red) overlaid on the photospheric image observed by the Helioseismic and Magnetic Imager (HMI) aboard the Solar Dynamic Observatory (SDO). Also shown is a VLA 3.8 GHz image (magenta) of the radio emission from the flare site. \textbf{b}, Closer view of the sunspot region (box in \textbf{a}). The 1.0 GHz and 3.8 GHz are shown as yellow and magenta contours, respectively, at 50\%, 70\%, and 90\% of the maximum}.
\label{fig:context}
\end{figure}

\begin{figure}[!htb]
\centering
\includegraphics[width=0.8\linewidth]{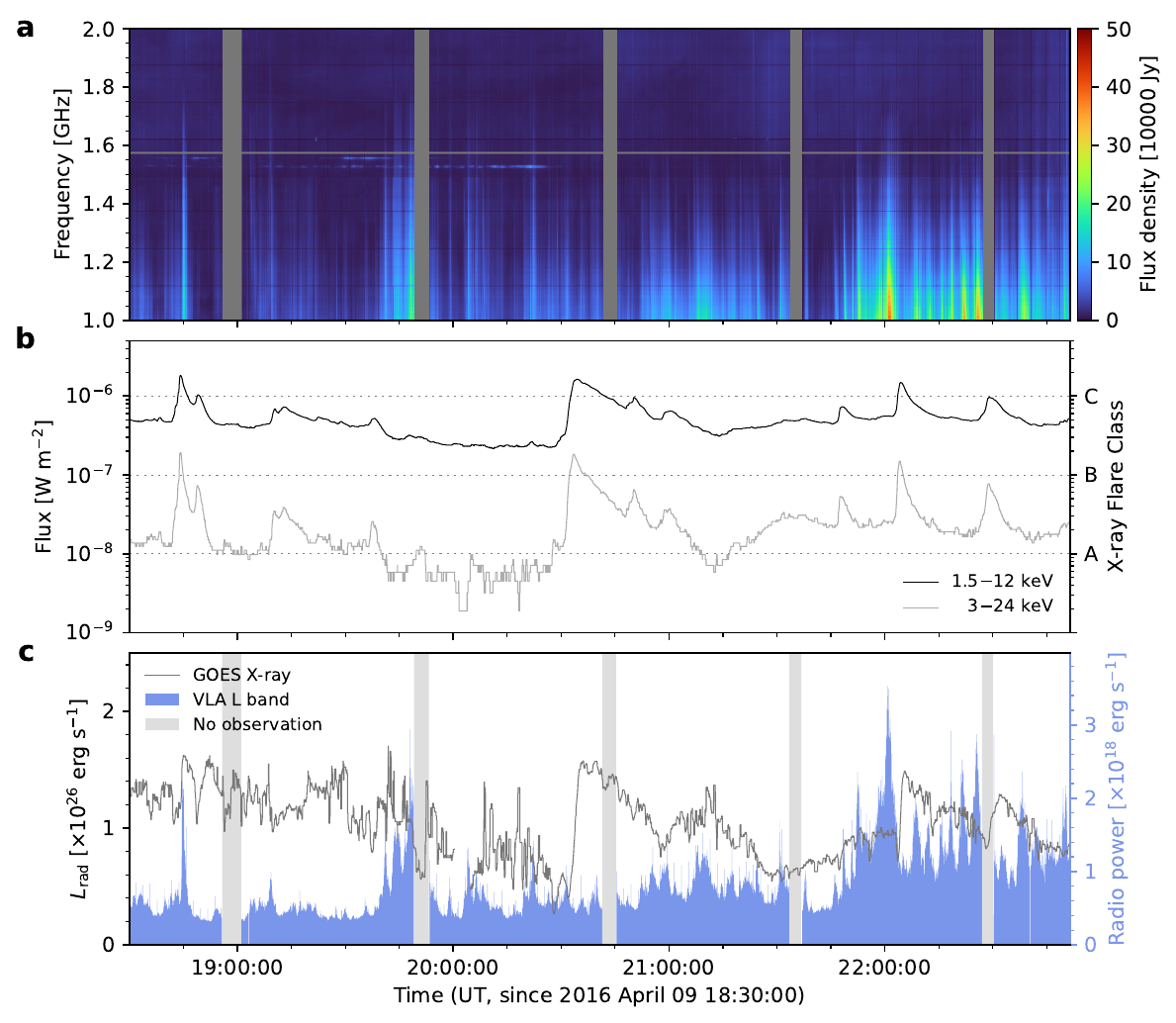}
\caption{\textbf{The relationship between X-ray and radio emissions.} \textbf{a}, Dynamic spectra of the right circularly polarized radio emission detected from the sunspot, as detected by the VLA. \textbf{b}, Time profiles of GOES 1.5--12 keV (black line) and 3--24 keV X-ray (gray line) X-ray, with corresponding GOES X-ray flare classes indicated. \textbf{c}, Time profile of the total radiative energy loss and the in-band (1--2 GHz) radio power for an isotropic emitter. Grey shaded bars represent gaps in the radio observations.}
\label{fig:time-seq}
\end{figure}

We observed the Sun using the Karl G. Jansky Very Large Array (VLA) on 2016 April 9 in both the L band (1--2 GHz) and the S band (2--4 GHz) using two sub-arrays. In the 1--2 GHz band, we detected a radio emission event from a bipolar sunspot group, dominated by a large leading spot with negative polarity (Fig. \ref{fig:context}). The dynamic spectrum of the radio source (Fig. \ref{fig:time-seq}a) obtained by the VLA reveals long-duration, repeated bursts across a broad frequency range from 1 to 1.7 GHz, persisting for the entire 4.5--h observation, with distinct bursts occurring at an average rate of $\sim$12 $\mathrm{h^{-1}}$. These bursts are also found in simultaneous 1 GHz solar radio flux data collected by the Nobeyama Radio Polarimeters (NoRP) and extend down to 245 MHz over an entire week when data from the Radio Solar Telescope Network (RSTN) is available (Extended Data Fig. \ref{fig:supp_norp_rstn}\&\ref{fig:supp_radio_profile}), indicating their broadband nature ($\delta\nu/\nu>150\%$). The radio bursts appear ``auroral-like'', as their temporal, spectral, and polarization characteristics, including high circular polarization fraction and brightness temperature, wide bandwidth, and extended duration, resemble the documented planetary aurora radio emissions and stellar radio ECM emissions.

\begin{figure}[!htb]
% \begin{mdframed}[leftmargin=5pt, rightmargin=5pt, linecolor=black!50, backgroundcolor=red!10, linewidth=2pt]
\centering
\includegraphics[width=0.9\linewidth]{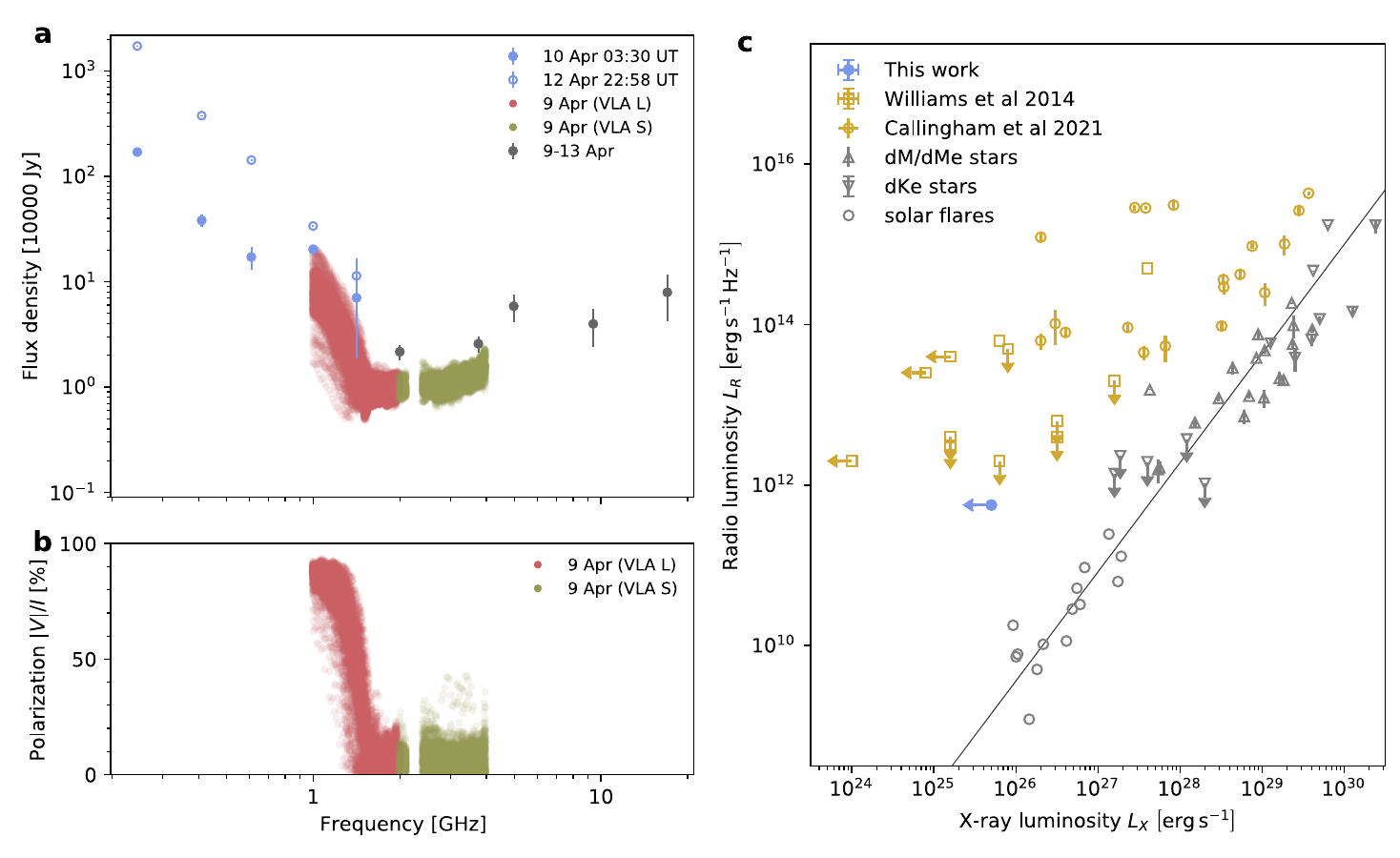}
% \includegraphics[width=0.45\linewidth]{fig/Guedel_Plot_v2.pdf}
% \end{mdframed}
\caption{\textbf{Synoptic radio spectra of the NOAA 12529 AR.} \textbf{a}, Stokes $I$ spectra detected with VLA, RSTN and NoRP measured between April 9--13, 2016. The peak spectra of the highly variable burst component measured by RSTN and NoRP at frequencies below 2 GHz on April 10 and 12 are shown as filled and open blue circles with error bars, respectively. 
% \SYS{A best-fit power-law function applied to the April 12 spectrum is represented by the dotted line.}
The spectra at frequencies above 2 GHz during April 9--13, dominated by incoherent gyrosynchrotron emission, are shown as gray circles. \SYS{The error bars represent the $5\sigma_{rms}$ and $1\sigma_{rms}$ uncertainties at each frequency for frequencies below and above 2 GHz, respectively}. \textbf{b}, The polarization degree (Stokes $V/I$) of the radio emission from the active region detected with VLA as a function of frequency. \textbf{c}, G\"{u}del--Benz (GB) relation between radio luminosity $L_R$ and X-ray luminosity $L_R$ for a wide range of magnetic active stars (gray symbols). Stellar radio emissions at frequencies of a few GHz and hundreds of MHz are represented as yellow squares and circles, respectively. The sunspot radio emission in this study is denoted by the blue circle.}
\label{fig:spec}
\end{figure}

The sunspot group, part of NOAA active region (AR) 12529, is 52$^\circ$ east and 10$^\circ$ north of the solar disk center. The radio source, $\sim40''$ (30 Mm) from the polarity inversion line (Extended Data Fig. \ref{fig:supp1}), coincides with sporadic flare activity observed by Geostationary Operational Environmental Satellites (GOES) in X-ray and the Atmospheric Imaging Assembly (AIA\cite{2012SoPh..275...17L}) onboard the Solar Dynamics Observatory (SDO) in (extreme) ultraviolet ((E)UV). The peak flux density of the radio source in stokes \textit{I} is approximately 20 sfu (1\,sfu=$10^4$\,Jy) at 1\,GHz and 2000 sfu at 245\,MHz (Fig. \ref{fig:spec}(a)), with an equivalent radiation brightness temperature of $10^{11}\,\mathrm{K}$ at 1\,GHz and $10^{12}\,\mathrm{K}$ at 245\,MHz, based on an inferred source size of 1,000\,km (Method). The brightness temperature is several orders of magnitude greater than the thermal temperature of the flaring plasma in the active region $\mathrm{T}\approx 6\times10^6\,\mathrm{K}$, constrained by the emission measure analysis based on the GOES X-ray data (Methods). Notably, the temporal variation of the radio luminosity from the sunspot source is poorly correlated with the X-ray luminosity of the flares (see Fig. \ref{fig:time-seq} and Extended Data Fig. \ref{fig:supp_norp_rstn}\&\ref{fig:supp_radio_profile}), suggesting that the bulk of the radio emission is not directly generated by the thermal and nonthermal electron population trapped in the flare sites.

The overall radio flux density spectrum of the active region (Fig. \ref{fig:spec}(a)) consists of two distinct spectral components, exhibiting a minimum near 2 GHz (Methods). We attribute the radio emission above the frequency minimum to incoherent gyrosynchrotron radiation emitted by mildly relativistic electrons energized by the flare process in the AR. The 1-minute VLA snapshot at 3.8 GHz obtained at 21:52 UT shows a compact source near the bright EUV flare arcade (Fig. \ref{fig:context}) with a relatively low degree of circular polarization $\lesssim20\%$ (Fig. \ref{fig:spec}(b)). Its source location is consistent with microwave imaging spectroscopy observations of other flares at similar frequencies, which have been interpreted as gyrosynchrotron radiation \cite{2020NatAs...4.1140C}. Given the proximity of the source to the flare arcade, low polarization degree, and flux density, we conclude that the radio emission at frequencies above 2 GHz is mainly gyrosynchrotron emission.

% Using the observed X-ray flux, we predict the expected radio flux density according to the G\"{u}del--Benz (GB) radio--X-ray relation\cite{1994A&A...285..621B}, which describes a tight relation between the observed X-ray and radio intensity in both solar flares and coronal emission from magnetically active stars (methods). The derived values at 5 GHz are consistent with the actual observed values (Fig. \ref{fig:spec}(a)). 

\begin{figure}[!htb]
% \begin{mdframed}[leftmargin=5pt, rightmargin=5pt, linecolor=black!50, backgroundcolor=red!10, linewidth=2pt]
\centering
\includegraphics[width=1\linewidth]{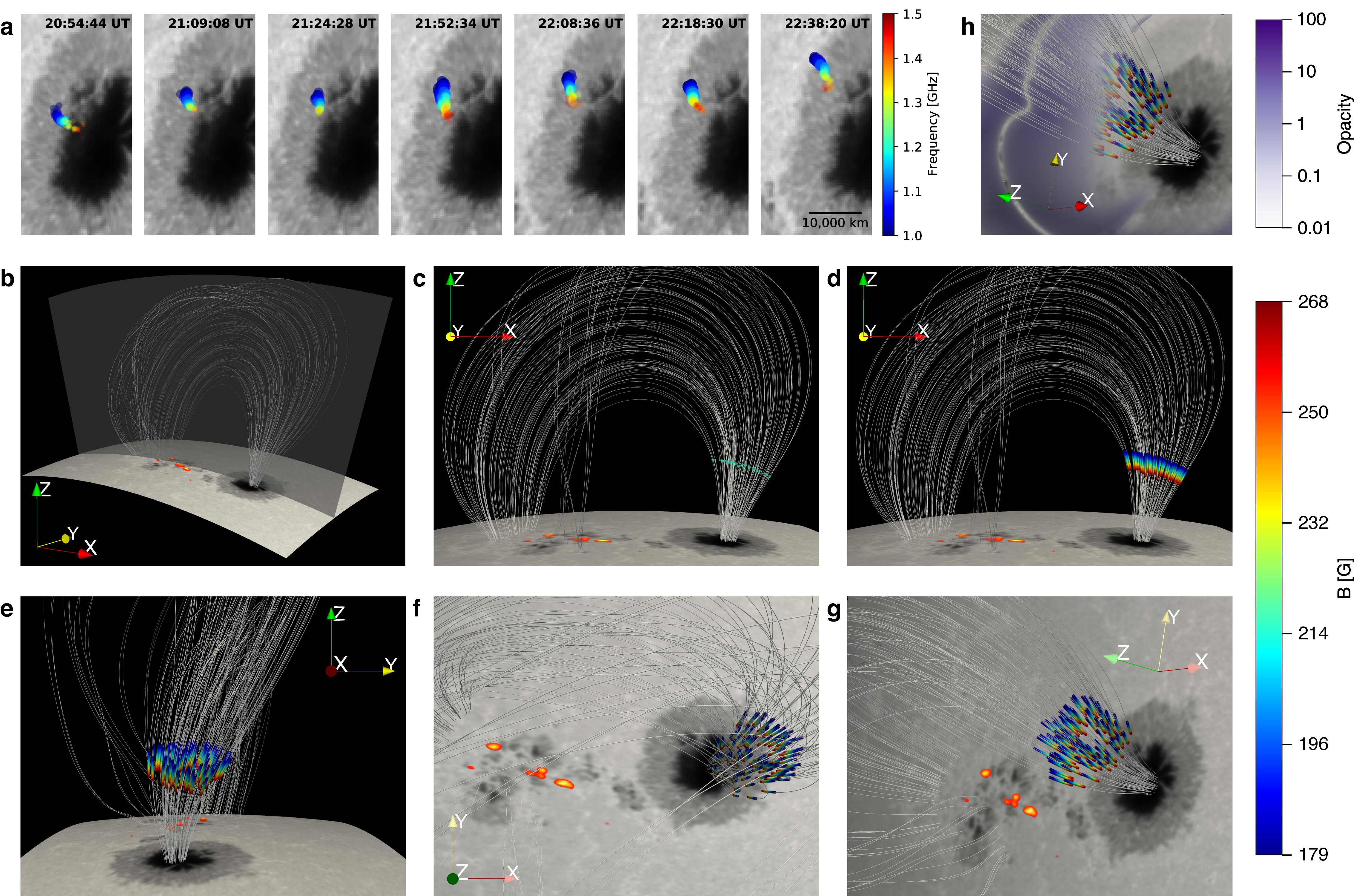}
% \end{mdframed}
\caption{\textbf{3D modeling of NOAA 12529 AR and the associated ECM emission sites.} \textbf{a}, Detailed view of the sunspot auroral radio source (white box in Fig. \ref{fig:context}b) at seven example times. The spatial distribution of the ECM sources appears as the convergent magnetic field lines about the sunspot is clearly delineated by the ECM sources of different burst events. The grayscale background is the SDO/HMI continuum image of the photosphere. \textbf{b}, 3D magnetic field model of the AR, with $X$- and $Y$-axes pointing to heliographic longitude and latitude on the solar disk, respectively, and $Z$-axis pointing radially from the Sun's center. White curves represent selected field lines for the closed fields connected to the sunspot umbra. Similar to Fig. \ref{fig:context}, reddish sources on the photospheric surface represent flare loop footpoints observed by AIA at 1600 \AA. \textbf{c}, Side view of the magnetic field model viewed towards positive $Y$-axis. The green dots above the sunspot denote the sites of the second harmonic ECM emission sources at $\nu_\mathrm{GHz} =1.2$ at selected field lines, outlining an iso-gauss dome of $B_\mathrm{G}=214$. \textbf{d}, Same as (c), but showing the ECM emission sites spanning a magnetic field strength from 179 to 268 G, colored according to field strength $B_\mathrm{G}$, which directly maps to ECM frequency $\nu_\mathrm{GHz}$ from 1.0 to 1.5\,GHz using the relation $B_\mathrm{G}\approx357\nu_\mathrm{GHz}/s$, where $s=2$ for the second harmonic ECM emission. \textbf{e--g}, Same as (d), but viewed towards negative $X$-axis, negative $Z$-axis, and the LOS, respectively. \textbf{h}, Same as (g), but overlaid with gyro-resonance opacity. Colorbars of magnetic field strength and opacity shown in (\textbf{c--h}) are displayed alongside.}
\label{fig:Bmodel}
\end{figure}

For frequencies below 2 GHz, the emission is dominated by the highly variable burst component. From 1.7 GHz to 245 MHz, the emission spectrum shows an increase of three orders of magnitude in flux density toward lower frequencies (Fig. \ref{fig:spec}(a)). The emission is nearly $100\%$ right-hand circularly polarized in the frequencies between 1.0 to 1.4\,GHz (Fig. \ref{fig:spec}(b)). Moreover, the flux density of the highly variable burst component does not follow the G\"{u}del--Benz (GB) radio/X-ray relation, which describes a tight relationship between the observed X-ray and radio intensity from solar and stellar coronae\cite{1994A&A...285..621B} (Methods). Specifically, the ratio of the radio to X-ray luminosities for frequencies below 1\,GHz violates the GB relation by up to more than two orders of magnitude (Fig. \ref{fig:spec}(c)). Using dynamic imaging spectroscopy with VLA, we have made images at hundreds of frequencies from 1 to 1.5 GHz at numerous selected times. During burst periods, all radio images are dominated by a localized radio source near the sunspot, exhibiting clear dispersion in space at different frequencies. We derive the source centroid of each radio image made at a given frequency. In Fig. \ref{fig:Bmodel}(a), we show the spatial distribution of the centroid locations by frequency for seven example time intervals, each spanning a duration of 20 seconds. The frequency-dependent source centroid locations of each burst form a close-to-linear dispersion in space, with high-frequency ends oriented towards the sunspot. Despite the overall location of the radio source relative to the sunspot remaining relatively stationary throughout the entire duration of the burst group, the distribution of centroid locations of individual bursts occurring at different times displays a rapid variation in the frequency-spatial dispersion and an evident spread in their position angles regard to the sunspot (Fig. \ref{fig:Bmodel}(h)). 
% In the following, we argue that the spatial distribution of the source locations of the radio bursts at different frequencies and times is the signature of radio sources aligned with the different converging magnetic field lines rooted at the sunspot.

\begin{figure}[!htb]
\centering
\includegraphics[width=1.0\linewidth]{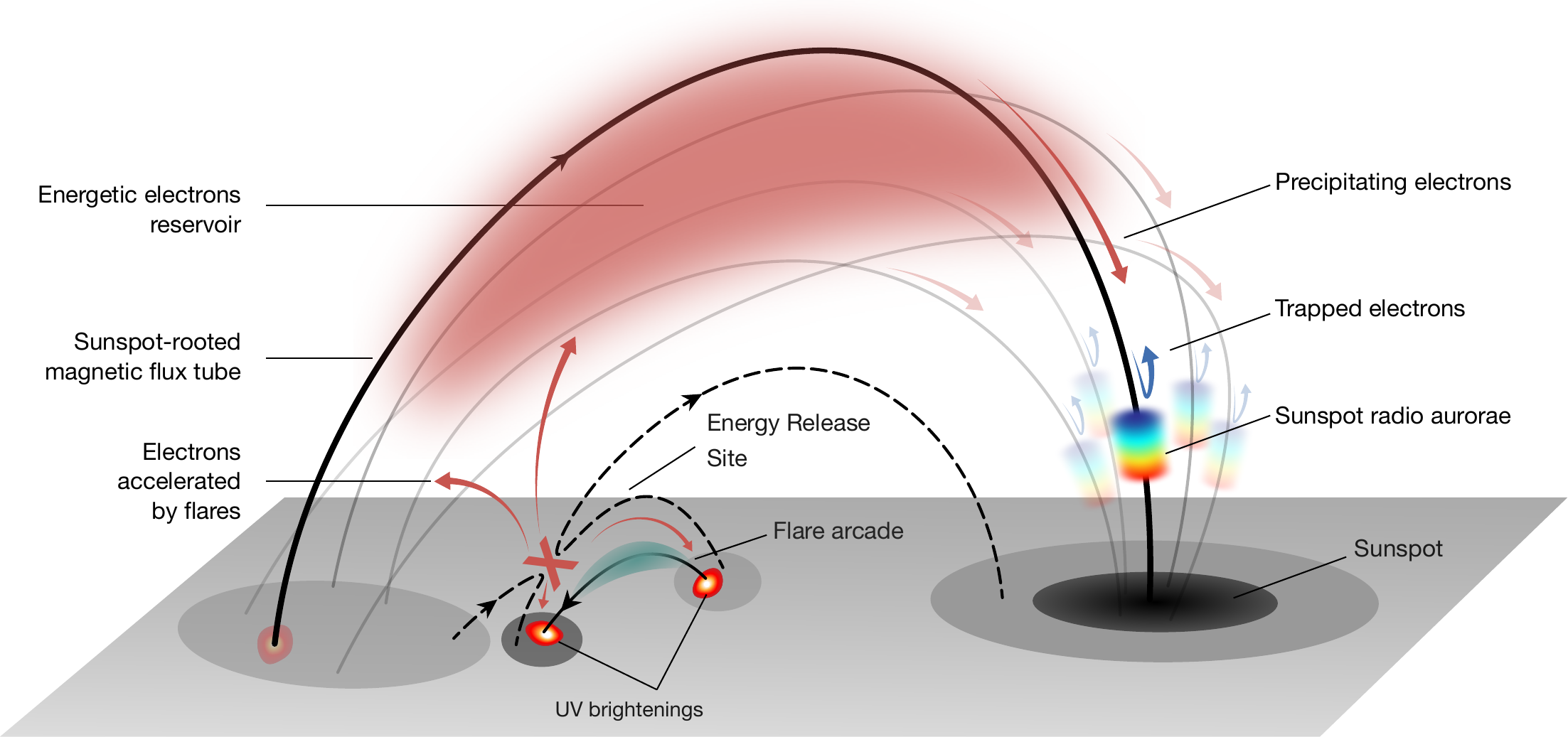}
\caption{\textbf{Schematic sketch of the sunspot auroral radio emission.} Physical scenario of electron acceleration by flaring activity, resulting in sunspot radio aurorae. Individual sunspot-rooted magnetic field lines filled with trapped electrons are responsible for individual auroral radio emission events.}
\label{fig:explain}
\end{figure}

The high circular polarization degree and brightness temperature both confirm that the observed emission is coherent in nature. We employ data-constrained coronal magnetic field and plasma density modeling 
% , based on magnetic field measurements made at the photosphere and multi-band EUV imaging data, respectively, 
to investigate the physical properties of the emission site (Methods). We find a region above the sunspot with a ratio of the electron plasma frequency $\nu_\mathrm{pe}$ and the electron gyrofrequency $\nu_\mathrm{ce}$ smaller than unity ($R=\nu_\mathrm{pe}/\nu_\mathrm{ce}<1$; Extended Data Fig. \ref{fig:params}(c)).
% , due to the presence of the sunspot. 
 % (3500 G at the photosphere)
% due to strong gyromagnetic absorption of the fundamental emission by the second harmonic absorbing layer (see Methods for details). 

The distribution of the predicted radio source locations in frequency (Fig. \ref{fig:Bmodel}(h)) aligns well with the spectro-spatial characteristics of the observed radio sources (Fig. \ref{fig:explain}a). The near-linear spatial dispersion as a function of frequency is well reproduced by individual unresolved emission sites aligned with the magnetic field lines. The magnetic field strength increase towards the sunspot accounts for higher-frequency sources (emitted from higher $B_{\rm G}$ regions) being closer to the sunspot. The overall spectro-spatial distribution of the source location of all the bursts with different position angles can be explained by the projection of distinct radio-hosting magnetic field lines into the plane of the sky (Extended Data Fig. \ref{fig:dis_freq} and Methods). 

Our analysis suggests most radio-hosting field lines have inclination angles of $30^{\circ}$ to $60^{\circ}$ relative to the line of sight (Extended Data Fig. \ref{fig:params}(f)\&\ref{fig:dis_freq}), consistent with the possible range for second-harmonic ECM emission to escape through a transparent window at the overlying absorbing layer at the third harmonic (Extended Data Fig. \ref{fig:params}(f)). We note that the visibility of the observed radio emission is likely a geometric coincidence of the beaming angle, the source location, and the transparent opacity window, which are subjected to change with time due to solar rotation and temporal evolution of the active region. A consequence of the rather stringent visibility requirement is that the radio emissions are likely only visible for a small fraction of the time over the entire course of the presence of the active region on the solar disk. The small filling factor in time may also explain why the sunspot radio ECM emission is observed only when the sunspot is between longitudes of $-70^\circ$ to $-10^\circ$ (east hemisphere) and sharply diminishes afterward, despite an increased flare activity during the sunspot's west hemisphere transit (Extended Data Fig. \ref{fig:supp_radio_profile}; Methods).

% unstable to ECM emission
\section*{Discussion}
Our interpretation of the observation is illustrated in Fig. \ref{fig:explain}(c). The observed coherent radio emission is interpreted as ECM emission induced by persistent energetic electrons trapped in the large-scale magnetic loops anchored at the sunspot. Analogous to the case of planetary radio aurora emissions\cite{2006A&ARv..13..229T}, the converging magnetic field above the sunspot can form a low $R$ cavity, providing a favorable geometry for radio ECM emission. Energetic electrons in the loops with initial pitch angles exceeding a threshold reflect at the magnetic mirror on the sunspot end of the loops, developing an anisotropic velocity distribution unstable to ECM emission in the cavity.
% In this scenario, the converging magnetic loops above the sunspot provides a favorable geometry---similar to the case of a low $R$ cavity in the planetary auroral case\cite{2006A&ARv..13..229T}---for the incoming energetic electrons to be trapped and develop a velocity distribution unstable to ECM emission. 
The source energetic electrons are likely associated with the sporadic flare activity occurring in the nearby active region, which is responsible for accelerating electrons to high energies and supplying them into the overarching magnetic loop system. The persistent radio emission suggests a continuous, albeit sporadic, operation of maser radiation, necessitating frequent replenishment of energetic electrons to maintain the electron anisotropy in the radio source. We speculate that weak yet frequent flare activity in the AR without apparent X-ray enhancement may drive electron acceleration\cite{2018ApJ...866...62C,2021ApJ...922..134B}.
%,2022ApJ...937...99S}. 
% This argument is supported by recent radio imaging spectroscopic observations of solar flare events across various scales, showing electron acceleration can occur even without notable activity in the X-ray flux\cite{2018ApJ...866...62C,2021ApJ...922..134B,2022ApJ...937...99S}. 

The absence of spatial coincidence and tight temporal correlation between flare activity and the coherent radio emission implies that the ECM emission is not directly produced by energetic electrons at the flare site.
% Instead, we suggest that the flare-accelerated energetic electrons are trapped in a ``reservoir'' in the overarching magnetic loops, which serves as the electron source for powering the observed long-duration ECM radio emission. 
Instead, the flare-accelerated electrons may have entered and become trapped in the overarching magnetic loops before precipitating to the ECM source region. During this process, transport effects such as collision, scattering, and trapping in the large-scale loops can act as an electron ``reservoir'' decoupling the time correlation between the flare activities and radio ECM bursts above the sunspot. %, likely determined by electron trapping time in the overarching loops. 
At the same time, the energetic electrons can be scattered into the loss cone and enter the region above the sunspot, driving and modulating the ECM emission.
We also note that magnetic field perturbations in the loop system, potentially caused by propagating disturbances originating from the sunspot\cite{1982Natur.297..485T,2023NatAs.tmp..110Y} or the flare site\cite{2018ApJ...864L..24L} with a period of 3--5 minutes, may affect the strength and direction of the magnetic field within the source region. Since ECM emission heavily relies on the local magnetic field, these perturbations could result in additional modulation in the observed temporal variability of the ECM emission. 
% The temporal variability of the coherent radio emission in a time scale of 2--5 minutes can be explained by perturbations of the magnetic loop system due to propagating disturbances from the sunspot\cite{1982Natur.297..485T} and/or the flare site\cite{2018ApJ...864L..24L} that moderating the electron precipitation into the sunspot and the subsequent ECM emission. 

The in-band radio power for an isotropic emitter, integrated over the L band (1--2 GHz), increases of up to 3.5 $\times 10^{18}\mathrm{erg}\,\mathrm{s}^{-1}$ during peaks against the pre-burst baseline level of $\sim$0.5 $\times 10^{18}\mathrm{erg}\,\mathrm{s}^{-1}$ (Fig. \ref{fig:time-seq}(c)). If the radio  emission at frequencies below 1\,GHz follows the spectral shape (Fig. \ref{fig:spec}(a)) recorded on April 12---the most powerful bursts recorded in the solar rotational cycle---the total radio power would be 100 times higher, reaching $10^{20}\mathrm{erg}\,\mathrm{s}^{-1}$. The flares release a total radiative energy of an order of $10^{26}\, \mathrm{ergs\ s^{-1}}$ as determined by X-ray diagnostics (Fig. \ref{fig:time-seq}(c)). Solar flares observations suggest that a significant fraction of flare energy is initially converted into non-thermal electron kinetic energy and subsequently transformed into various energy forms\cite{2017LRSP...14....2B}. Previous studies of solar energetic electrons observed at 1 AU suggest 0.01--1\% of the flare-accelerated energetic electrons escape from the energy release sites and enter the interplanetary space\cite{2007ApJ...663L.109K}. Using this fractional value as the lower limit to estimate the total power of the energetic electrons trapped in the corona, we obtain an average power of $>\!10^{22}$--$10^{24}\,\mathrm{ergs\,s^{-1}}$, more than two orders of magnitude greater than that of the radio emission. Therefore, we conclude that the flare-supplied energetic electrons should contain adequate energy to power the observed coherent radio emission.

% [electrons can be accelerated and escape from the corona even during a period without a major eruption. type III bursts observed by Lofar and PSP \url{https://ui.adsabs.harvard.edu/abs/2022arXiv220408497B/abstract} 
% The intensity of the radio emission shows no clear dependence on the X-ray luminosity.  ]

% The spectral characteristics of the observed radio emission, its persistent nature, and its violation of the radio/X-ray correlation, resemble, to some extent, the ECM emissions detected from stars spanning a broad spectral types from magnetic massive stars\cite{2011ApJ...739L..10T,2018MNRAS.474L..61D,2022ApJ...925..125D} to cooler objects such as M and brown dwarfs\cite{2007ApJ...663L..25H,2020NatAs...4..577V,2021NatAs...5.1233C}. Whereas ECM emissions from these objects are generally explained within paradigm of planetary auroral radio emission from a dipolar magnetosphere, our observations point to an alternative avenue to produce auroral radio emissions in flare stars via starspot-flare interaction. Those stars are notable for retaining vigorous and recurring magnetic activities in their coronae\cite{2008AJ....135..785W,2016ApJ...820...89F} that can power the required persistent electron-refilling. As the case presented here, electron acceleration be driven by local magnetic activities, even during a period without major flare eruptions. These together suggest that the flare-sunspot mechanism may operates well into the regime of sun-like stars and cooler objects.

Our findings \BC{may have important} implications for the interpretation of similar radio emissions observed at other stars and star-planetary systems\cite{2019ApJ...871..214V}\BC{, because they share similar spectral-temporal properties and the deviation of the GB radio/X-ray correlation \cite{2007ApJ...663L..25H,2020NatAs...4..577V,2021NatAs...5.1233C}.} 
% We note that the spectral-temporal properties and the apparent violation of the GB radio/X-ray correlation of our sunspot ECM emission observed at frequencies below 2\,GHz closely resemble radio emissions reported from M dwarfs\cite{2007ApJ...663L..25H,2020NatAs...4..577V,2021NatAs...5.1233C}. We acknowledge the flux density of the observed bursts (2$\times10^{10}$ mJy, or 2000 sfu; Extended Data Fig. \ref{fig:supp_radio_profile}(d)) at 245 MHz, equivalent to a $\mu$Jy radio source located at a stellar distance of 20 parsecs, is three orders of magnitude lower than the mean values from recent results of LOFAR observations of M-dwarfs at hundreds of MHz \cite{2021NatAs...5.1233C}. 
%\SYS{While acknowledging that the flux density of the observed bursts (2$\times10^{10}$ mJy, or 2000 sfu; Extended Data Fig. \ref{fig:supp_radio_profile}(d)) at 245 MHz, which is equivalent to a $\mu$Jy radio source located at a stellar distance of 20 parsecs, is three orders of magnitude lower than the mean values from recent results of LOFAR observations of M-dwarfs at hundreds of MHz \cite{2021NatAs...5.1233C}, it's critical to note that the spectral-temporal properties and the apparent violation of the GB radio/X-ray correlation of the sunspot ECM emission closely resemble radio emissions reported from M dwarfs\cite{2007ApJ...663L..25H,2020NatAs...4..577V,2021NatAs...5.1233C}.} 
%We suggest that the discrepancy in the radiation intensity is likely due to the relatively low magnetic field strength of the sunspot and the low magnetic activity level in the active region (Methods).
Stellar ECM emissions are generally explained within the paradigm of planetary auroral radio emission from a dipolar magnetosphere, driven by either star-planet interactions\cite{2020NatAs...4..577V} or co-rotation breakdown of a plasma disk\cite{2017MNRAS.470.4274T} {for cooler dwarf stars, and centrifugal breakout reconnection for rapid-rotating early type stars\cite{2022MNRAS.513.1449O}}.
% , and reconnection at the magnetic equator for early-type massive stars\cite{2004A&A...418..593T}. 
These stars feature magnetic activity\cite{2008AJ....135..785W,2016ApJ...820...89F} and are notable for the presence of starspots\cite{2020ApJ...897..125R}. If these stars have a continuously replenished population of energetic electrons due to flaring activity, these electrons would have the opportunity to be trapped above a starspot and facilitate the growth of ECM emission, even without a globally dipolar magnetic field geometry. This scenario may also be viable for interpreting certain radio emissions from the apparently non-flaring, quiescent stars\cite{2005ApJ...627..960B,2020NatAs...4..577V}, \SYS{which show a similar magnitude of deviation from the tight GB relation in their radio/X-ray correlation}: a large number of small-scale energy release events may be at work to supply energetic electrons sufficient for powering a persistent radio source under favorable emission conditions. This scenario is akin to our case in which many episodes of intense radio bursts are observed during the quiescent period when there is no flare activity. However, we note that the interpretation may not be applied toward the ECM radio aurorae on the coolest ultracool dwarfs\cite{2015ApJ...815...64W,2016ApJ...818...24K,2016MNRAS.457.1224L,2017ApJ...846...75P,2018ApJS..237...25K}, which lack starspots with a stable loop and flaring energetic electron acceleration. \BC{We also note that our observed bursts above the sunspot are three orders of magnitude weaker than those reported on certain flare-active M-dwarfs with extremely high brightness temperatures (up to 10$^{14}$ K for those reported by, e.g., \cite{2021NatAs...5.1233C}; see Methods for more discussions). While such a difference in radiation intensity is an outstanding topic for further investigation, we suggest that the extremely high magnetic field strength and activity level on these M dwarfs could be a contributing factor.} %\SYS{We also note that, while this model provides a plausible explanation for mJy level stellar ECM emissions, extending its applicability to interpret Jy level emissions\cite{2021A&A...648A..13C} poses a substantial challenge (Method).}

Periodic radio emission occasionally observed from stars, coinciding with stellar rotation, has been interpreted as emanating from their polar regions\cite{2005ApJ...627..960B,2009ApJ...695..310B,2007ApJ...663L..25H,2008ApJ...684..644H,2015Natur.523..568H,2017ApJ...836L..30L,2018MNRAS.474L..61D}, similar to planetary auroral emission. Given the presence of a persistent starspot lasting for multiple rotations \cite{2020ApJ...897..125R} and recurring flaring activity supplying energetic electrons, we argue that the sharply-beamed ECM emission above the starspots can also lead to auroral-like radio bursts with the rotation period of the host star. It is noted that the emission may also be modulated by the short- and long-term variability of the flare activity of the host star, which may be responsible for the observed variations in the burst duty cycle across different timescales detected from some stars \cite{2006ApJ...653..690H,2007ApJ...663L..25H,2008ApJ...673.1080B,2008ApJ...676.1307B,2011ApJ...739L..10T} (Methods). Previously observed stellar ECM emissions may not be interpreted exclusively as magnetospherically-driven auroral emissions from the polar regions, but could also originate from regions above strong starspots. In sum, our results not only clarify the origin of a type of radio burst on the Sun, but also prompt a re-examination of the physical mechanisms behind certain auroral-like radio emissions from other stars. This study also highlights the need for more detailed observations in the future to better understand the intricacies of the ECM emission on the Sun, including the responsible plasma instability, local driver of the ECM (e.g., a parallel electric field\cite{2000ApJ...538..456E}), and energy efficiency of the ECM emission, which remains largely unexplored.

%TC:ignore

\section*{Methods}

\renewcommand\thefigure{\arabic{figure}}
\renewcommand{\figurename}{Extended Data Fig.}
\setcounter{figure}{0}

% publicly available at NOAA National Geophysical Data Center (NGDC) Website \url{http://www.ngdc.noaa.gov} and Australian Government Bureau of Meteorology World Data Centre (WDC) Website \url{https://www.sws.bom.gov.au/World_Data_Centre}. 

\subsection*{GOES X-ray data reduction}
\begin{figure}[!htb]
% \begin{mdframed}[leftmargin=5pt, rightmargin=5pt, linecolor=black!50, backgroundcolor=red!10, linewidth=2pt]
\centering
\includegraphics[width=0.95\linewidth]{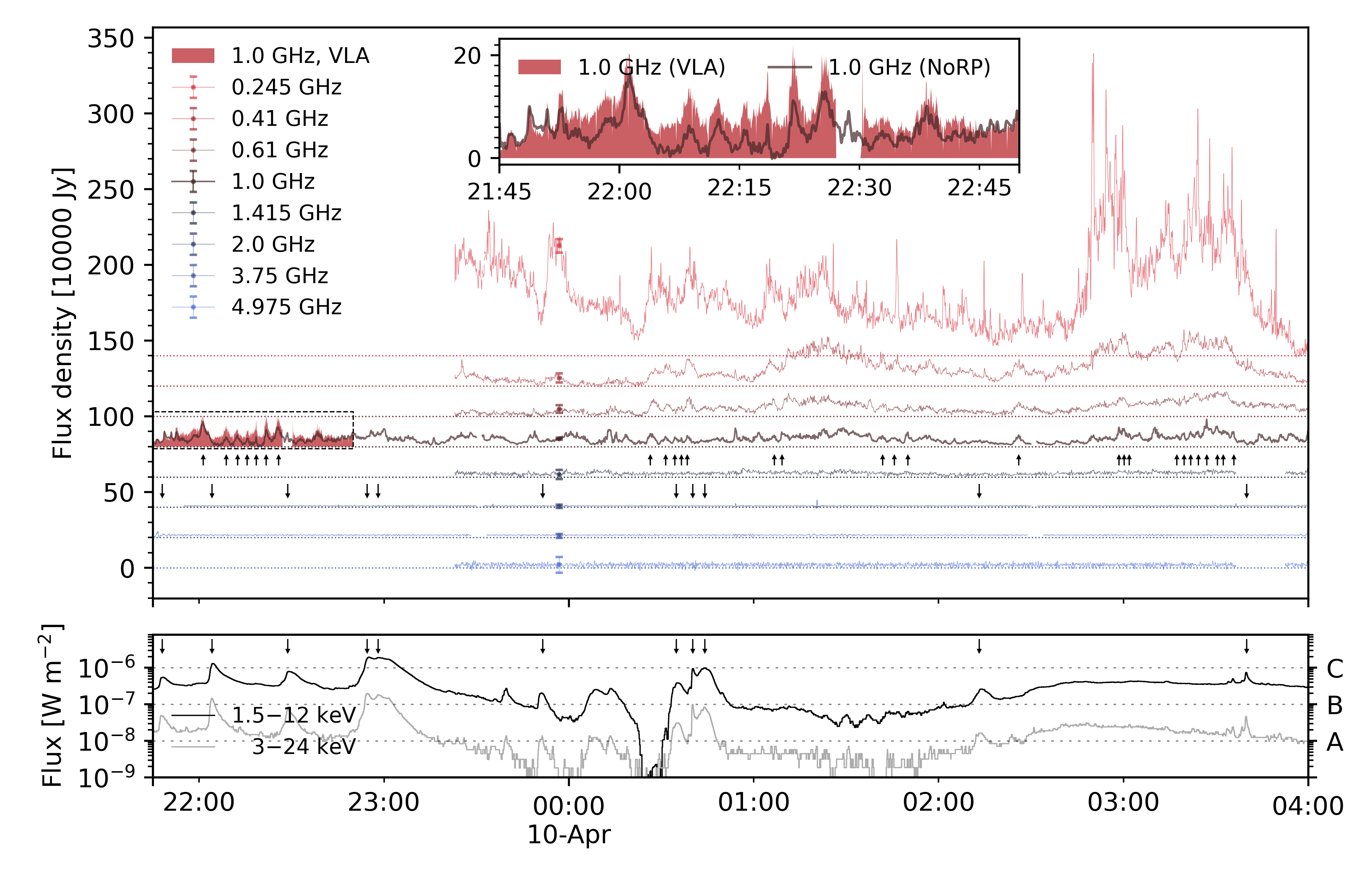}
% \end{mdframed}
\caption{\textbf{Radio and X-ray flux of the radio-burst-hosting active region (NOAA 12529 AR) observed by RSTN, NoRP, and GOES on 2016 April 9--10.} \textbf{Top panel}: The background subtracted solar radio flux at discrete frequencies monitored by RSTN, NoRP and VLA. The horizontal dotted lines in corresponding colors indicate equally spaced zero-lines for each frequency channel, set at 20$\times10^4\,\mathrm{Jy}$ (or 20 sfu) intervals. The error bars overlaid on the zero-flux-lines in corresponding colors show the $3\sigma_{rms}$ uncertainties of each frequency channel. The upward arrows indicate the most prominent examples of sunspot radio aurorae, while the downward arrows indicate the occurrence of flare events. The inset shows a blowup of the 1 GHz flux simultaneously observed by NoRP and VLA, denoted by the dashed box. No zero-line offset was applied for the inset.
\textbf{Bottom panel}: Simultaneous time profiles of GOES 1.5--12 keV (black line) and 3--24 keV X-ray (grey line), with corresponding GOES X-ray flare classes indicated. The arrows are the same as the downward arrows in the top panel.}
\label{fig:supp_norp_rstn}
\end{figure}

The X-ray Sensor (XRS) onboard the Geostationary Operational Environment Satellite (GOES) continuously monitors the visible solar hemisphere in two energy bands, one in 1--8 \AA\ (or 1.5--12 keV) and another in 0.5--4 \AA\ (or 3--24 keV), both of which are dominated by thermal continuum emission. The GOES/XRS data were reduced using the standard analysis tool available in the SolarSoftWare IDL ({\tt SSWIDL}) package (\url{https://www.lmsal.com/solarsoft/ssw_setup.html}). Because AR 12529 is the only flaring AR at the time of the radio observation, we attributed the impulsive X-ray emission shown in Extended Data Fig. \ref{fig:supp_norp_rstn}(bottom) to thermal Bremsstrahlung emitted by the coronal plasma heated by the flare activity in this AR. We subtracted a constant background from the XRS time series, and compute the total radiative energy loss rate shown in Fig. \ref{fig:time-seq}(c) from the expressions \cite{2005SoPh..227..231W} available in the {\tt SSWIDL GOES} tool. We note The mean square root flux uncertainty of the 1.5--12 keV and 3--24 keV channels is $5\times10^{-9}$ and $2\times10^{-10}$ $\mathrm{W\,m^{-2}}$, respectively, each of which is two orders of magnitude lower than the corresponding background level. Additionally, since the flares exceed the background level by one order of magnitude at their peak, the detailed choice of the background for subtraction does not impact the results.

\subsection*{Radio data reduction: VLA}
The observation was made when VLA is in C configuration, for which the longest antenna baseline is over 3 km, yielding an angular resolution of $17''.3\times29''.3/\nu^2_\mathrm{GHz}$ at the time of the observation. We observed the Sun using two subarrays in dual-polarization mode at 50 ms cadence, with 512 and 1024 frequency channels of 2 MHz spectral resolution, covering 1--2 GHz (L band; 14 antennas) and 2--4 GHz (S band; 13 antennas), respectively. The flux calibration, bandpass calibration, and complex gain calibration are performed using a standard celestial calibrator 3C48. When observing the Sun, 20 dB attenuators were switched into the signal path of each antenna and each polarization. The delay and amplitude introduced by the attenuators were removed following the technique described in Chen et al (2013) \cite{2013PhDT.......498C}. Data were reduced using the \textsl{Common Astronomy Software Applications} (CASA Release 5.4; \url{http://casa.nrao.edu}).

\begin{figure}[!htb]

% \begin{mdframed}[leftmargin=5pt, rightmargin=5pt, linecolor=black!50, backgroundcolor=red!10, linewidth=2pt]
\centering
\includegraphics[width=0.8\linewidth]{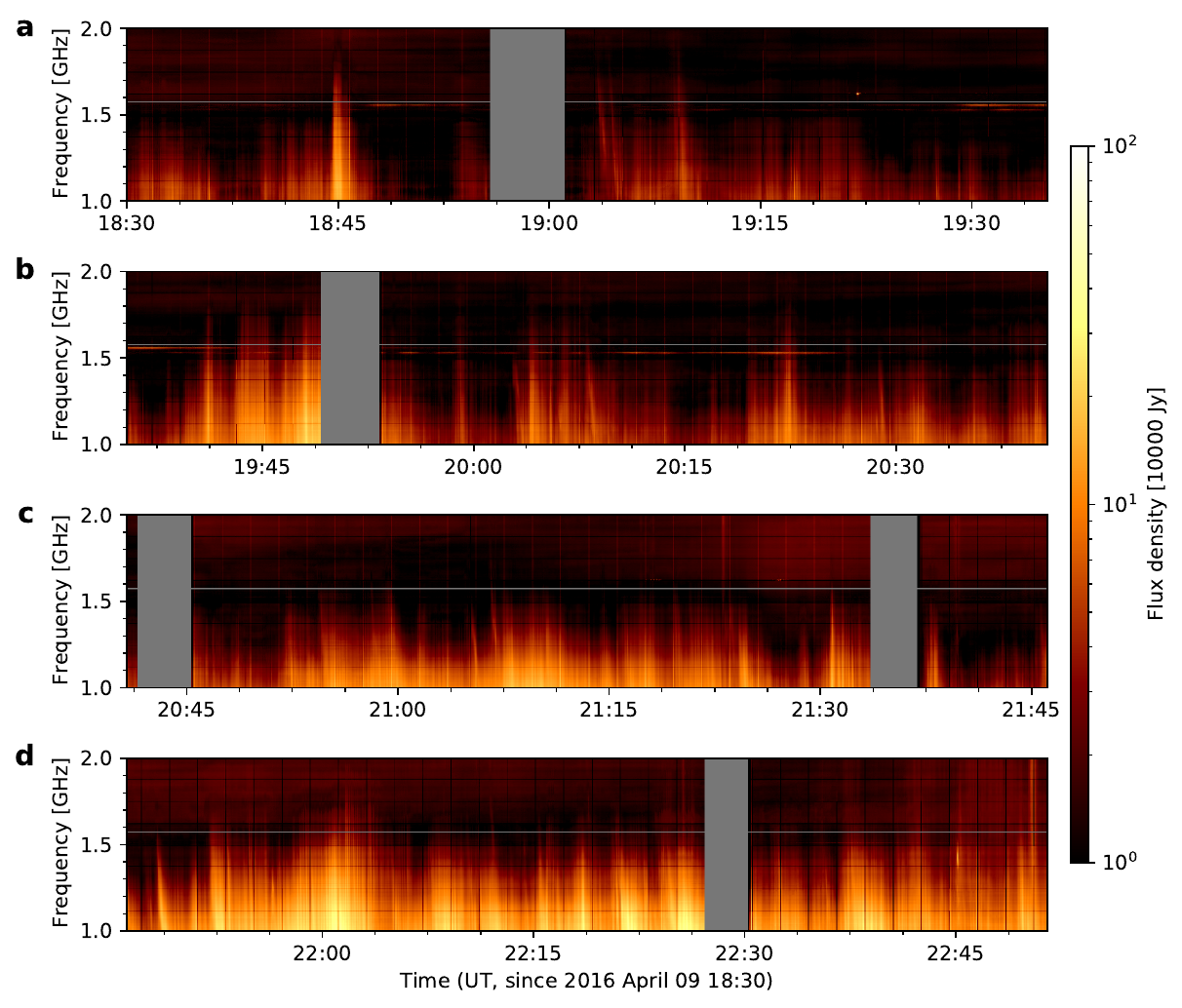}
% \end{mdframed}
\caption{\textbf{Dynamic spectra of the right circularly polarized radio emission detected from the sunspot with VLA on April 9, 2016.}}
\label{fig:VLA_dspec}
\end{figure}

\begin{figure}[!htb]
\centering
\includegraphics[width=0.4\linewidth]{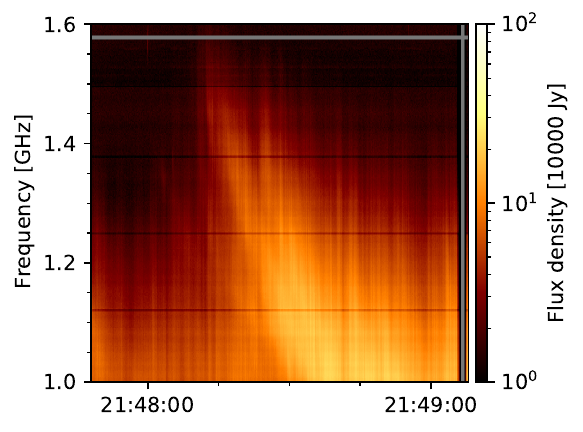}
\caption{\textbf{Details of the dynamic spectrum between 21:48 and 21:49 UT on April 9, 2016.}}
\label{fig:VLA_dspec_details}
\end{figure}

The data contaminated by radio frequency interference, particularly those near GPS L1 frequency (1.58 GHz), are flagged. The slow-varying background emission contributed by the quiescent Sun is subtracted from the VLA data. The spectrotemporal intensity variation (dynamic spectrum) intrinsic to the sunspot radio source are estimated by averaging the amplitude of the complex visibility data over short baselines (0.2--0.8 km) that are sensitive to the spatial scale of the active region. The radio source is located at an angular distance of $13'.4$ from the antenna pointing at solar disk center, which is comparable to the size of the primary beam of the VLA antennas $\theta_{\rm PB}=45'/\nu_{\rm GHz}$ in the frequency range of 1--4 GHz, where $\nu_{\rm GHz}$ is observing frequency in GHz. To ensure the absolute flux density of the source is correct, we scale the observed flux density of the VLA dynamic spectrum at L and S band up to that at the antenna pointing center according to the primary beam response (a Gaussian beam pattern is used). An overview of the radio emission from the sunspot in the L band is shown in Extended Data Fig. \ref{fig:VLA_dspec} for the 4.5 hr duration of the observation. Extended Data Fig. \ref{fig:VLA_dspec_details} shows a detail of the L band dynamic spectrum of 1 minute observation after 21:48 UT.

Radio imaging of each pixel in the dynamic spectrum where the bursts are found provides key information on the spatial variation of the radio source as a function of time and frequency. We produce independent radio images at all the spectral channels between 1.0 to 1.5 GHz for every four time pixels in the dynamic spectrum using the CASA task \texttt{tclean}. We note a periodic pattern is present in the observed visibility phases as a function of time. This is due to the telescope not updating its geometric delays at a rate that matches the time-cadence of the data, known as the so-called ``delay clunking'', which introduced an offset with regard to the true phase center a fixed rate of $-2''.5\mathrm{/min}$ in right ascension. This effect is mitigated by inserting an offset of $2''.5\mathrm{/min}$ along the same direction to the sky coordinates of the radio images. It is known that the absolute position of the source can be determined to a small fraction ($\sim10\%$) of the synthesized beam under typical conditions, due to atmospheric phase stability, the SNR on target, etc. We verified the alignment accuracy of the radio images (without background subtraction) using features in the NOAA 12529 active region seen in both radio and EUV wavelengths. Since the source of the sunspot radio bursts is only marginally resolved by VLA, we treat the radio source above 50\% of the maximum as a point-like source. The uncertainty of the centroid location of the source $\sigma$ can be determined by\cite{1997PASP..109..166C}:
\begin{equation}
    \sigma \approx \theta_\mathrm{HPBW}/(\mathrm{SNR}\sqrt{8\ln{2}})\ ,
\end{equation}
where $\theta_\mathrm{HPBW}$ is the half-power beamwidth of the synthesized beam and SNR is the signal-to-noise ratio of the image. With a typical SNR value for the bursts exceeding 100, we can infer a position accuracy to $\lesssim 0''.2$. However, the propagation of radio waves through the inhomogeneous corona toward the observers is known to be susceptible to scattering effects\cite{1994ApJ...426..774B,2017NatCo...8.1515K}. Therefore, the estimate of uncertainty given above should only be considered as a lower limit. In fact, by obtaining the centroid locations of all frequency–time pixels on the sunspot radio bursts within a short time period (40\,s) and frequency range (40\,MHz), we find that they are distributed rather randomly within an area of FWHM size of $\sim1''.5\times1''.5$ (Extended Data Fig. \ref{fig:dis_freq}). Hence, we estimate the actual position uncertainty of the centroids as $\sigma\approx1''.0$.

\begin{figure}[!htb]
% \begin{mdframed}[leftmargin=5pt, rightmargin=5pt, linecolor=black!50, backgroundcolor=red!10, linewidth=2pt]
\centering
\includegraphics[width=\linewidth]{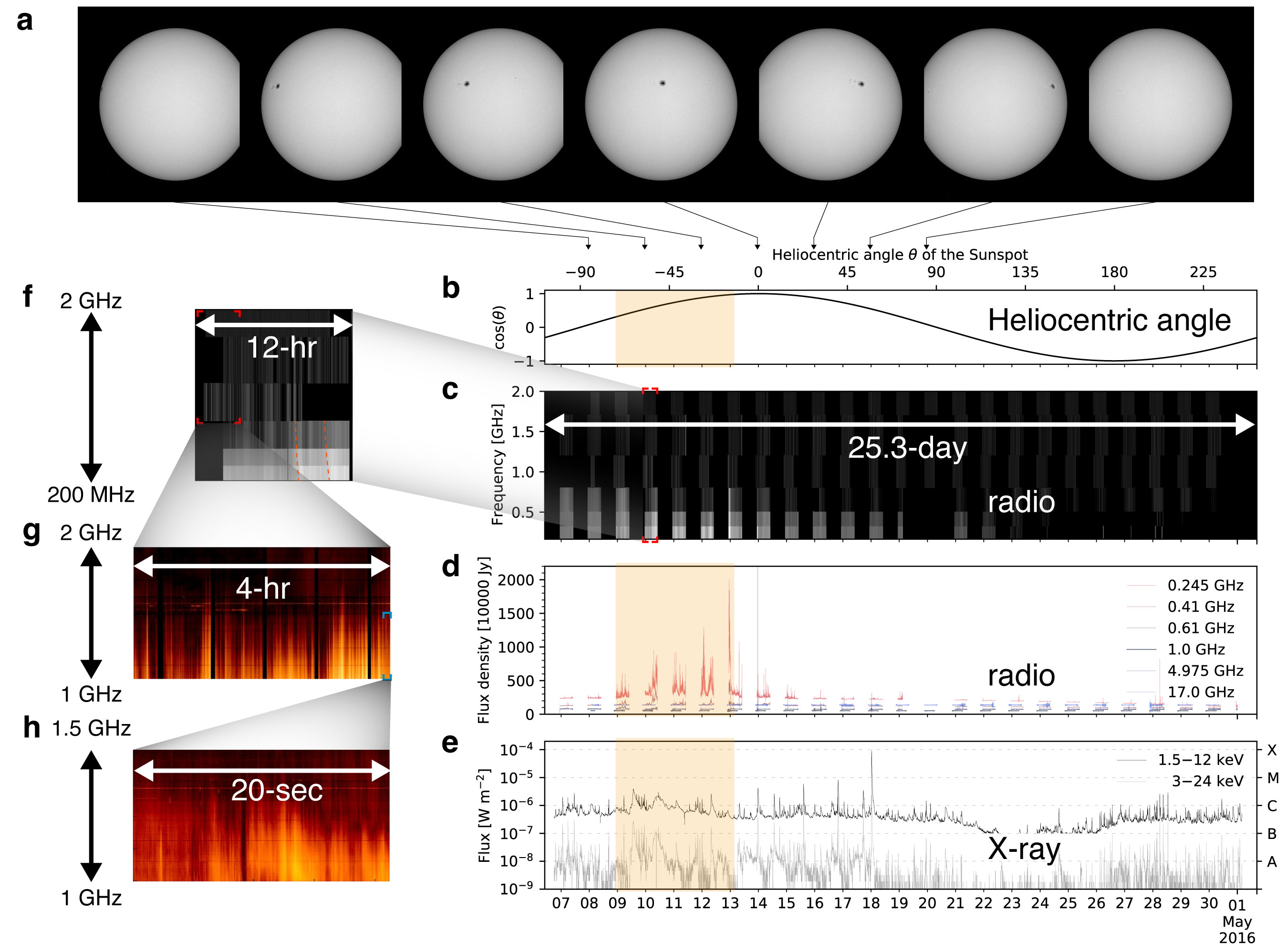}
% \end{mdframed}
\caption{\textbf{Flux profiles of radio-hosting active region (NOAA 12529 AR) over one solar rotation cycle.} \textbf{a}, SDO/HMI continuum image sequence showing the transiting sunspot at times marked by the vertical arrows in (b). \textbf{b--e}, Time history of the heliocentric angle $\theta$ of the sunspot, the radio, and X-ray emission of the Sun over one solar rotation. \textbf{f}, Blowup of the synoptic dynamic spectrum of Stokes I obtained by NoRP and RSTN in (c), showing the arcs (red dashed lines) of radio emission. \textbf{g--h}, Hierarchical substructures of the emission arcs shown in (f) obtained by VLA.}
\label{fig:supp_radio_profile}
\end{figure}

\begin{figure}[!htb]
% \begin{mdframed}[leftmargin=5pt, rightmargin=5pt, linecolor=black!50, backgroundcolor=red!10, linewidth=2pt]
\centering
\includegraphics[width=\linewidth]{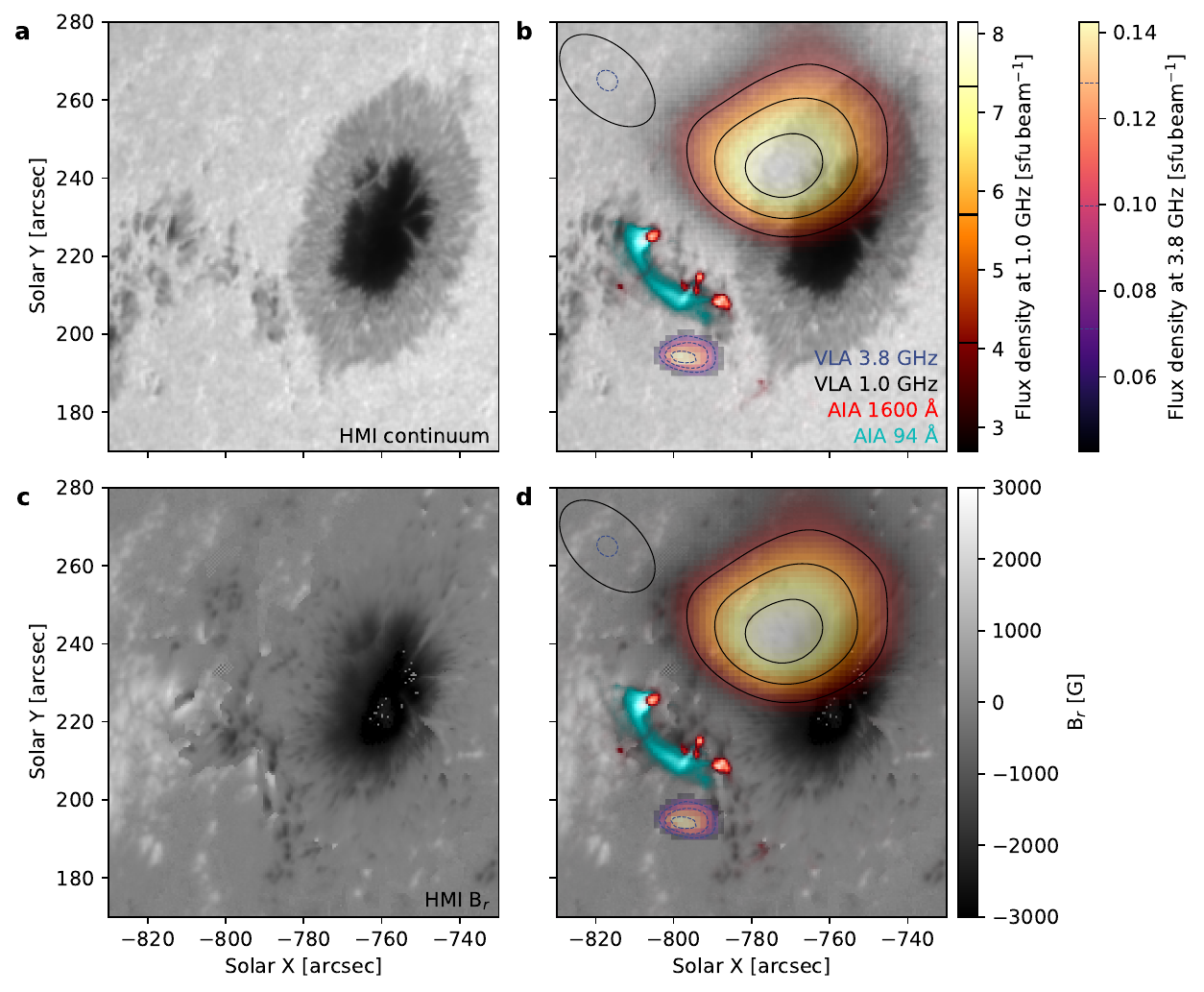}
% \end{mdframed}
\caption{\textbf{Photospheric images of NOAA 12529 AR.} \textbf{a--b}, Continuum intensity image at 6173 \AA\ at the solar surface, obtained by the HMI on board SDO. \textbf{c--d}, Radial component of the photospheric magnetic field strength. In the right column, the coherent (L-band) and incoherent (S-band) radio sources are shown as yellow and purple, respectively. Contours for the two radio sources at 50\%, 70\%, and 90\% are overlaid as black solid and blue dashed lines. The size of the synthesized beams for the L and S band images are displayed in the upper-left corner with the corresponding line styles. The L- and S-band radio images were made using VLA observation on 2016 April 9 at 22:08 and 21:53 UT, respectively. The flare loops and their footprints, as observed by SDO/AIA 94 and 1600 \AA, are shown in blue and red, respectively.}
\label{fig:supp1}
\end{figure}

\subsection*{Radio data reduction: RSTN and NoRP}
USAF Radio Solar Telescope Network (RSTN) monitors the sun-as-a-star radio flux at 8 fixed frequencies --- 0.245, 0.41, 0.61, 1.415, 2.695, 4.995, 8.8 and 15.4 GHz --- at a cadence of one second. The data in use were recorded by the RSTN Learmonth site. The daily radio fluxes at frequency $\nu_{\rm GHz}> 1$ show an overall inverted-U shape, which is mainly due to the varying zenith-angle-dependent opacity owing to atmospheric absorption of the radio waves. To correct for this effect, we first fit the day-to-day variation of the local noon radio flux at each frequency by using a third-order polynomial function over one solar rotation cycle from 2016 April 6 to 2016 May 2. Next, the daily inverted-U shape of the radio light curves in each frequency due to atmospheric absorption were fit by using an eighth-order polynomial function and subsequently removed. Finally, the resulting light curves are scaled to the fitted day-to-day noon flux curve. In addition, the Nobeyama Radio Polarimeters \cite{1985PASJ...37..163N} (NoRP) perform full-stokes polarimetric observations of the Sun at six frequencies --- 1, 2, 3.75, 9.4, 17, 35 GHz --- with a temporal resolution of 0.1 second. Data was reduced using the {\tt NoRP} package in the SolarSoftWare IDL package (\url{https://lmsal.com/solarsoft/ssw/radio/norh/}). We used a similar data processing method to correct the daily shape presented in the NoRP data. 

In the absence of flares, the gradually varying background presented in the reduced RSTN and NoRP radio data at frequencies above 2 GHz is contributed mainly by the quiescent active region and the solar disk due to thermal free-free and gyroresonance emission. Since the background variations have a time scale that is much longer than that of flares, a high-pass filter was then used to remove the slow-varying components from the RSTN and NoRP data. To determine the flux uncertainty of each frequency channel, we calculated the root mean square ($\sigma_{rms}$) of the temporal fluctuations during two quiescent time intervals of 3--4 UT and 6--7 UT on April 8.

The Extended Data Fig. \ref{fig:supp_norp_rstn}(top) presents the resulting daily solar radio fluxes on April 9--10, 2016.  The frequency range between 1 to 2 GHz is dominated by the repetitive radio bursts with a typical cadence of 5--10 minutes, which are not well correlated with the X-ray flux, as depicted in Fig. \ref{fig:time-seq}a. The Extended Data Fig. \ref{fig:supp_norp_rstn}(top) and its inset panel show the background-subtracted data at 1 GHz obtained simultaneously by both the VLA and NoRP. The 1 GHz light curves from NoRP show a deviation no greater than 50\% from that obtained by VLA, indicating a high degree of agreement between the two datasets and further verifying the robustness of the background subtraction approach. Despite the VLA observation ending at 22:50 UT on 2016 April 9, the same emission feature observed by the VLA is still evident in the NoRP at 1 GHz after 22:50 UT. Furthermore, we show the fluxes below 1 GHz obtained by RSTN in Extended Data Fig. \ref{fig:supp_norp_rstn}(top). The burst component in the 1 GHz flux, denoted by the upward arrows, is also notably presented in fluxes at lower frequencies down to 0.245 GHz. The temporal characteristics of the light curves when the bursts occur are similar, indicating the broadband nature of the bursts that should extend to well below 1 GHz. This broadband nature aligns with the rising spectrum toward lower frequencies shown in the VLA data of 1--2 GHz (L band). However, it is worth noting that the RSTN data lacks polarization measurements. As a result, in our interpretation, we have heuristically assumed that the radio bursts observed in the non-VLA data at frequencies below the L band are circularly polarized coherent radio emissions. We made this assumption based on two facts: 1) the bursts observed by the VLA, NoRP, and RSTN have a similar temporal profile at distinct frequencies below 2\,GHz, and 2) the flux density of the bursts is too high for incoherent radio emission without major solar eruptive events.

Overall, the observations presented in Extended Data Fig. \ref{fig:supp_norp_rstn} demonstrate a consistent temporal behavior across a broad range of frequencies. The agreement between the VLA and NoRP data at 1 GHz, along with the detection of bursts at lower frequencies by the RSTN, provides compelling evidence for the broadband nature of these bursts. However, further studies incorporating polarization measurements and high-spectral resolution in a wider frequency range are necessary to gain a comprehensive understanding of these bursts and their polarization properties at different frequencies.

% \subsection*{Variability of the radio emission over a solar rotation}

Extended Data Fig. \ref{fig:supp_radio_profile} shows the time history of the solar radio and X-ray emission for an entire solar rotation period. The temporal and spectral variation of the radio emission indicates that the same coherent emission feature persists for over an entire week (or $\sim$1/4 of the rotation period; demarcated by the yellow shaded area in Extended Data Fig. \ref{fig:supp_radio_profile}(b, d--e)). The HMI continuum image series in Extended Data Fig. \ref{fig:supp_radio_profile}a show that AR 12529, which the sunspot belongs to, is the only AR on the disk when the sunspot radio emission is present, therefore we attributed the impulsive X-ray emission shown in Extended Data Fig. \ref{fig:supp_radio_profile}(e) to the flare activity in this AR. The GOES X-ray flux level displays sporadic flare activity during the time when the sunspot moves across the solar disk, and drops to the background level after the sunspot moves to the back side of the Sun. The radio emission shows little correlation with that of the X-ray emission from flare activity in the AR. In fact, the radio emission is only visible for a limited duration during the rotation cycle, specifically when the sunspot is located between longitudes of -70$^\circ$ to -10$^\circ$. The temporal, spectral, and polarization characteristics of the radio emission, including broadband arc-like structures in the time-frequency domain (Extended Data Fig. \ref{fig:supp_radio_profile}f), closely resemble ECM emissions reported in the literature from (sub)stellar systems\cite{2015Natur.523..568H,2022ApJ...935...99B}. Most notably, the sun-as-a-star light curves throughout a solar rotation, primarily modulated by the coherent sunspot radio bursts (Extended Data Fig. \ref{fig:supp_radio_profile}(c--d)), display a striking similarity to rotationally-modulated periodic radio aurora from M-dwarfs\cite{2019MNRAS.488..559Z}. The similarity is mainly evident in the temporal patterns in radio emissions. In both cases, the radio emissions exhibit rotationally modulated behavior, which corresponds to the rotation of the respective celestial bodies. Specifically, in the solar case, the burst duty cycle---the fraction of time spent in bursting---is approximately 20\% in the L and P bands. This value is comparable to those of coherent radio bursts observed in M-dwarfs within the same frequency range, as demonstrated in previously reported stellar instances.\cite{2019MNRAS.488..559Z,2019ApJ...871..214V}.
% Most notably, striking similarity is found between the sun-as-a-star light curves over a solar rotation, modulated mainly by the radio bursts (Extended Data Fig. \ref{fig:supp_radio_profile}(c--d)), and rotationally-modulated periodic radio aurora from M-dwarfs\cite{2019MNRAS.488..559Z}.

The radio bursts from the sunspot contribute significantly to the flux variations at frequencies below 2\,GHz, while at frequencies above 2\,GHz, the temporal fluctuations are mainly caused by incoherent gyrosynchrotron emission generated by energetic electrons resulting from solar flares. To obtain the RSTN and NoRP data points in the radio burst spectra shown in Fig. \ref{fig:spec}, two different approaches were adopted for frequencies below and above 2\,GHz, respectively. The upward arrows in Extended Data Fig. \ref{fig:supp_norp_rstn}(top) denote examples of sunspot radio bursts at 1\,GHz that occurred on April 9--10. We easily identified the burst component against the noise levels of the time series, and used the peak-to-trough amplitude of the most prominent burst that occurred at 03:30 UT on April 10 as the radio flux density of the corresponding frequencies. \SYS{Nonetheless, it is important to highlight that this method inherently eliminates any constant flux component of the sunspot radio emission. As such, the flux density at frequencies below 2 GHz should be interpreted as a lower limit only.} The resulting spectrum in the 1--2 GHz range deviates no more than $\pm$50\% from the maximum of the VLA background subtracted spectra obtained between 21:45 and 22:50 UT on April 9. We used the same method to obtain the spectrum of the strongest sunspot radio burst in this solar rotation cycle, which was recorded at 22:58 UT on April 12. The downward arrows in Extended Data Fig. \ref{fig:supp_norp_rstn} denote the X-ray flare events that occurred on April 9--10. For frequencies above 2\,GHz, the radio light curves of the flares are just above or even comparable to the noise levels. Therefore, we took the peak fluxes of individual frequencies in the entire time interval when the sunspot radio emission is present (April 9--13). However, we note that the low signal-to-noise ratio may introduce bias in the flux estimation for frequencies above 2\,GHz. Thus, the flux estimated at these frequencies should be considered as upper limits.

% 15\%

% \begin{figure}[!htb]
% \centering
% \includegraphics[width=0.75\linewidth]{fig-sxr_EM.png}
% \caption{}
% \label{fig:supp2}
% \end{figure

% , typically integrated over the 0.2--2.0 keV energy range, with the radio luminosity $L_R$ at 5 GHz, or alternative frequencies if 5 GHz data is unavailable. 
% The background level was subtracted by fitting a third-order polynomial to the pre- and post-flare values of each fare.

\subsection*{The radio--X-ray G\"{u}del--Benz relation}
The G\"{u}del--Benz (GB) empirical relation connects the radio and X-ray emission for a broad range of magnetically active stars, with the relationship $L_R \approx 10^{-15.5\pm0.5}~L_X~\mathrm{Hz}^{-1}$, where $L_R~\mathrm{[erg~s^{-1}~Hz^{-1}]}$ denotes radio luminosity and $L_X~\mathrm{[erg~s^{-1}]}$ represents X-ray luminosity within soft X-ray energy range \cite{1994A&A...285..621B}. In Fig. \ref{fig:spec}c, we present the best linear fit to the data from Benz \& G\"{u}del (1994)\cite{1994A&A...285..621B}, as given by Berger et al. (2010)\cite{2010ApJ...709..332B},
\begin{equation}
    L_R = 1.36 (L_X-18.97).
    % \mathrm{log}(L_R) = 1.36(\mathrm{log}(L_X)-18.97).
\label{equ:gbr}
\end{equation}
However, as discovered in the past two decades, this correlation may not hold true for certain late-type dwarf stars. The radio/X-ray correlation of stellar ECM radio emission, spanning a broad frequency range between hundreds of MHz and a few GHz as documented in the literature\cite{2010ApJ...709..332B,2014ApJ...785....9W,2021NatAs...5.1233C}, are evidently divergent from the GB relation. This deviation is exemplified by the yellow scatters depicted in Fig. \ref{fig:spec}c. 

\begin{figure}[!htb]
% \begin{mdframed}[leftmargin=5pt, rightmargin=5pt, linecolor=black!50, backgroundcolor=red!10, linewidth=2pt]
\centering
\includegraphics[width=0.45\linewidth]{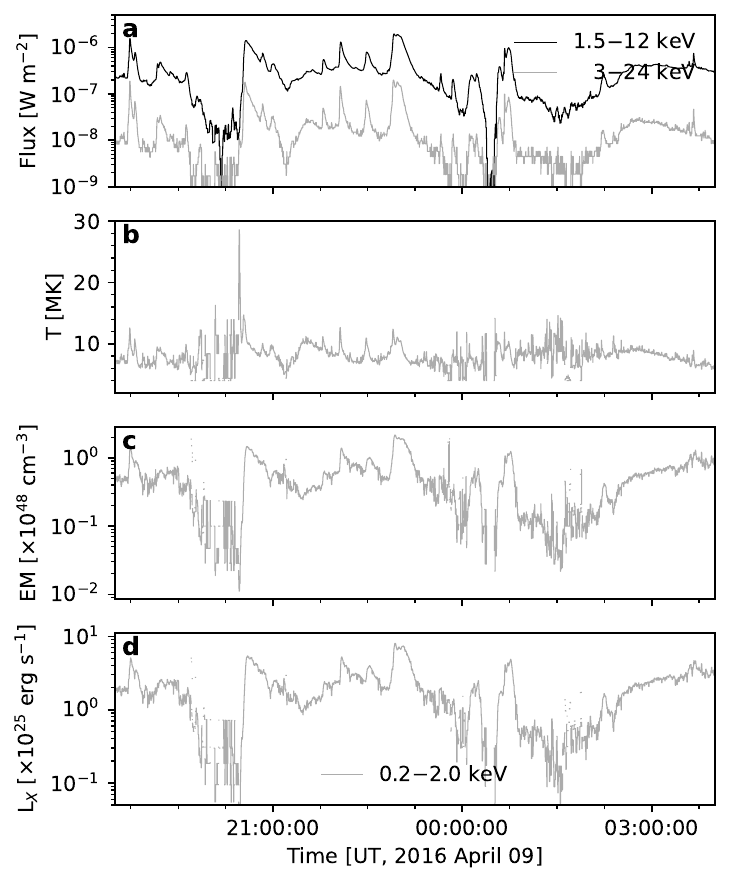}
\includegraphics[width=0.45\linewidth]{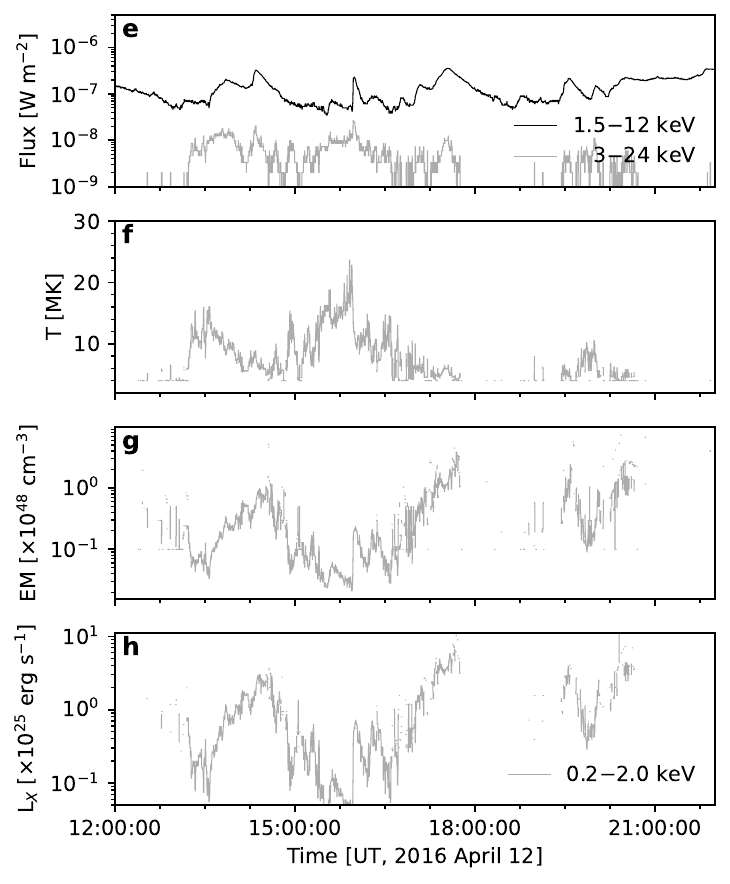}
% \end{mdframed}
\caption{\textbf{Temporal evolution of the flare activity in NOAA 12529 AR on April 9th (left) and 12th (right).} Displayed sequentially from top to bottom are GOES 1.5–12 keV (black line) and
3–24 keV X-ray (grey line), the derived plasma temperature $T$ and total emission measure $EM$ determined based on the background-subtracted X-ray emission, as well as the X-ray luminosity in 0.2--2.0 keV ($L_X$) employing the Raymond-Smith plasma model.}
\label{fig:sxr_EM}
\end{figure}

The simultaneous radio and X-ray observations allow us to assess the observed sunspot radio ECM bursts in the context of the GB relation. In order to draw a comparison between the sunspot radio luminosity $L_R$, which reached its peak at 245 MHz, and the solar X-ray $L_X$ luminosity, we calculate the $L_X$ using the following approach. The $L_X$ of the Sun at the time of the observation is dominated by the flare activity in NOAA AR 12529. We used the {\tt SSWIDL GOES} tool to model the background-subtracted X-ray emission and estimate the total emission measure $EM$ and a temperature $T$ with an isothermal model and photospheric elemental abundance. Subsequently, the calculated $EM$ and $T$ were integrated into the Raymond-Smith plasma model\cite{1977ApJS...35..419R} to determine the total $L_X$ in the energy band of 0.2--2.0 keV---a common used energy band in the literature\cite{2010ApJ...709..332B,2014ApJ...785....9W,2021NatAs...5.1233C} (Extended Data Fig. \ref{fig:sxr_EM}). The radio/X-ray correlation of the sunspot radio emission is shown as the blue circle in Fig. \ref{fig:spec}c, displaying a clear deviation from the GB relation by more than two orders of magnitude, comparable to the observations of stellar ECM radio emission documented in previous research\cite{2010ApJ...709..332B,2014ApJ...785....9W,2021NatAs...5.1233C}.

% With each $L_X$ value, we estimate the corresponding radio luminosity $L_R$ using the relation of Equation (\ref{equ:gbr}). 
% The resulting $L_R$ level of the flares is shown as the horizontal gray bar in Fig. \ref{fig:spec}(a).

\subsection*{Coronal magnetic field model}
The coronal magnetic field above the sunspot is not directly measured using, e.g., the Zeeman splitting technique, because the corona is extremely tenuous and optically thin. To reconstruct the coronal magnetic field in three dimensions, we perform a nonlinear force-free field (NLFFF) extrapolation based on measurements of vector magnetic field at the photosphere. More details of the assumptions and procedures for the NLFFF technique can be found in other works\cite{2004SoPh..219...87W}. In short, in the solar corona, the magnetic pressure dominates over non-magnetic forces. Therefore, under a quasi-static configuration (which is applicable to our case with a quasi-steady active region), the corona can be approximated as in a force-free state. In this case,
reconstructing the coronal magnetic field can be well described as a boundary value problem under the force-free field approximation\cite{1995ApJ...439..474M}. %The coronal magnetic field can be extrapolated from photospheric magnetic measurements. We used the weighted optimization NLFFF algorithm\cite{2004SoPh..219...87W} to reconstruct three-dimensional (3D) magnetic field of NOAA AR 12529 from the vector magnetogram in heliographic cylindrical equal-area projection,
The vector photospheric magentic field measurements were recorded by the Helioseismic and Magnetic Imager (HMI) instrument on board SDO \cite{2012SoPh..275..207S} at 22:00 UT, processed with the standard pipeline\cite{2014SoPh..289.3549B}. In addition, a pre-processing procedure was applied to reduce the net force and torque of the photospheric magnetic field\cite{2006SoPh..233..215W}.

%The vector photospheric magnetogram of the active region at 22:00 UT on 2016 April 9. 
The original photospheric vector magnetogram data has an angular resolution of $0.03\, ^{\circ}\,\mathrm{pixel}^{-1}$ in heliographic coordinate system\cite{2006A&A...449..791T}, which was binned to $0.06\, ^{\circ}\,\mathrm{pixel}^{-1}$ (equivalent to $\sim 720\,\mathrm{km}\,\mathrm{pixel}^{-1}$) for the subsequent extrapolation performed in a Cartesian grid of $400\times224\times224$. 

\subsection*{Coronal density model} \label{subsec:density model}
The corona density model above the sunspot is approximated using the plane-parallel atmosphere assumption. The coronal electron number density as a function of height $h$ above the solar surface $n_\mathrm{e}(h)$ is prescribed using the Baumbach-Allen formula originally derived from white-light observations during solar eclipses \cite{2000asqu.book.....C}:
\begin{equation}
    n_\mathrm{e}(h)=m\left[0.036 (\frac{h+R_\odot}{R_\odot})^{-1.5}+1.55 (\frac{h+R_\odot}{R_\odot})^{-6}+2.99 (\frac{h+R_\odot}{R_\odot})^{-16}\right] \times 10^8\,\mathrm{cm}^{-3},
\end{equation}
where $m$ is a factor to account for the enhanced overall density above the sunspot-hosting active region.

\begin{figure}[!htb]
% \begin{mdframed}[leftmargin=5pt, rightmargin=5pt, linecolor=black!50, backgroundcolor=red!10, linewidth=2pt]
\centering
\includegraphics[width=1.0\linewidth]{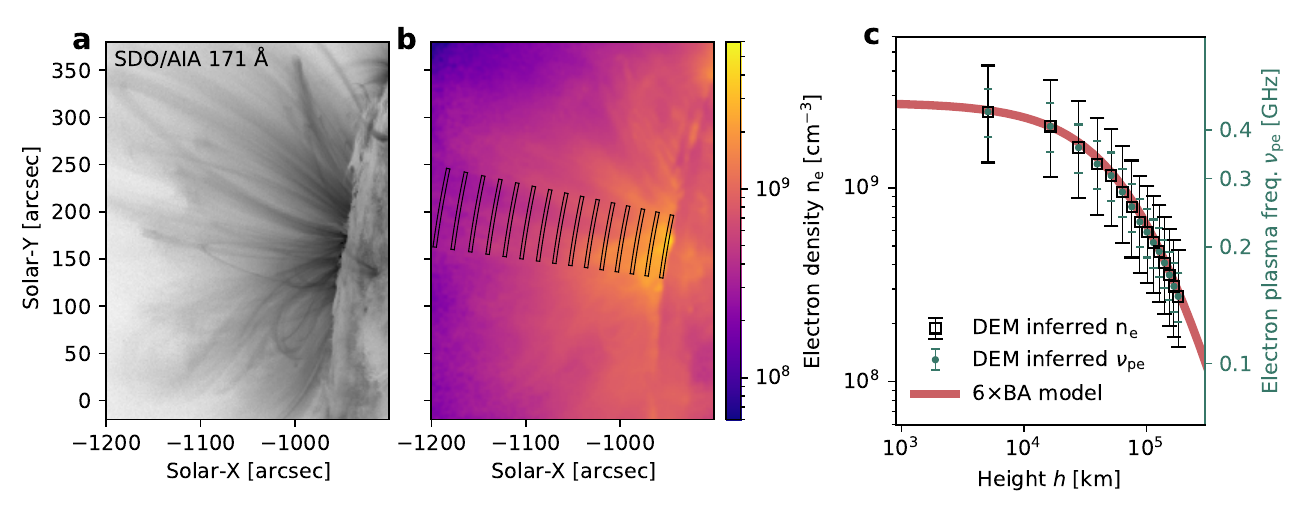}
% \end{mdframed}
\caption{\textbf{Electron number density distribution in the corona above the sunspot on 2016 April 7 when the sunspot was located at the east solar limb.} \textbf{a}, SDO/AIA 171 Å passband image at 08:10 UT. \textbf{b}, electron density $n_\mathrm{e}$ derived by EUV diagnostics. \textbf{c}, distribution of $n_\mathrm{e}$ and the corresponding electron plasma frequency $\nu_\mathrm{pe}$ in height $h$ obtained from regions marked in (b) for a series of equally spaced radial heights above the sunspot. The red line represents the sixfold Baumbach--Allen model. The error bars are obtained by with the column depth $D$ ranging from 15--150\,Mm.}
\label{fig:NEmodel}
\end{figure}

To determine an appropriate $m$ factor for our case, we note that the sunspot first appears on the east limb two days prior to the time of the events under study. The sunspot could be clearly distinguished as it rotated to its position on April 9 2016 from the east limb and does not show any significant morphological changes in the two-day period (Extended Data Fig. \ref{fig:supp_radio_profile}(a)), implying that the coronal magnetic and density structures above the sunspot should not vary significantly within the period. Thus, we further constrain the density model using multi-band EUV images obtained by SDO/AIA when the sunspot is located at the east limb. A differential emission measure (DEM) analysis is performed using the regularized inversion method\cite{2012A&A...539A.146H} based on imaging data at six SDO/AIA EUV bands (94~\AA, 131~\AA, 171~\AA, 193~\AA, 211~\AA, 335~\AA). The results of the analysis are a distribution of the line-of-sight-integrated total emission measure $\xi(x,y)=\int n_e^2(x,y)dz \approx n_e^2(x,y)\,D$ in the plane of the sky, where $D$ is the column depth weighted by the square of the density along the line of sight. In order to derive the density distribution $n_e(x,y)$ from the DEM results, we take a constant column depth of $D = 50\times10^3\,\mathrm{km}$. With the derived limb-view $n_e(x,y)$ map shown in Extended Data Fig. \ref{fig:NEmodel}(b), we can derive the $n_\mathrm{e}(h)$ distribution above the active region shown in Fig. \ref{fig:NEmodel}(c), in which each data point at $h$ above the limb is obtained by averaging the $n_\mathrm{e}$ values from the pixels enclosed in the black contours in Extended Data Fig. \ref{fig:NEmodel}(b). We find the sixfold Baumbach--Allen model is an excellent fit to the $n_\mathrm{e}(h)$ inferred from the DEM analysis. We caution, however, that the actual density distribution depends on the selection of the column depth along the line of sight, which is largely unknown. The column depth we adopted should only be considered as a rough order of magnitude estimation. Therefore, the derived $n_\mathrm{e}(h)$ distribution may vary by a scale factor of a few if we select a different column depth.

\subsection*{Emission mechanism of the sunspot radio bursts} \label{subsec:EmissionMechanism}
There are two major competing emission mechanisms that can generate coherent radio emissions in the solar corona. Coherent plasma radiation often dominates the solar radio emission of solar flares in a wide frequency range spanning from $\sim1$ GHz all the way to tens of MHz. They are driven by the non-linear growth of plasma instabilities and followed by a wave-wave conversion, radiating at a frequency close to the fundamental or second harmonic electron plasma frequency $\nu_\mathrm{pe}\approx 8980\sqrt{n_e}\ \mathrm{Hz}$. In contrast, the electron cyclotron maser (ECM) emission likely dominates the emission in regions of high magnetic field strength and/or low plasma density, where the electron plasma frequency ($\nu_\mathrm{pe}$) is less than the electron gyrofrequency ($\nu_\mathrm{ce}$) at the source site \cite{1982ApJ...259..844M}. The ECM mechanism has been suggested as the primary emission mechanism of auroral radio emission in planetary magnetospheres, such as the terrestrial auroral kilometric radiation \cite{1979ApJ...230..621W}, Jovian decametric radiation\cite{2007P&SS...55...89H}, and Saturn's kilometric radiation\cite{2008JGRA..113.7201L}, as well as strong radio radiation with high circular polarization in a variety of magnetically active stars such as M dwarfs \cite{2005ApJ...627..960B,2007ApJ...663L..25H,2015Natur.523..568H}. ECM emission is generated near low harmonics of the electron cyclotron frequency $s\nu_\mathrm{ce}=2.8\times10^6 sB\,\mathrm{Hz}$ (where $s$ is the harmonic number and $B$ is the magnetic field strength in Gauss) due to non-linear wave growth under a variety of plasma instabilities including loss-cone, ring shell, or horseshoe instabilities\cite{2006A&ARv..13..229T}. We recognize that details of the ECM model, including the responsible plasma instability, local driver of ECM (e.g., a parallel electric field, as suggested by Ergun et al. 2000\cite{2000ApJ...538..456E}), and energy efficiency of the ECM emission remain largely unexplored. Addressing these details requires further, and preferably, \textit{in situ} measurements to characterize the local magnetic and electric field, electron momentum and angular distribution, and other plasma parameters at the radio emitter\cite{2000ApJ...538..456E}.
% ECM emission is only possible in regions where the condition $\nu_\mathrm{pe}<s\nu_\mathrm{ce}$ is satisfied. 
% Since the electron number density $n_e$ and magnetic field strengths $B$ generally decrease with increasing height within a magnetic loop in the corona, and so do the plasma frequency $\nu_\mathrm{pe}$ and cyclotron frequency $\nu_\mathrm{ce}$, coherent radio emission at a higher frequency is expected to be located closer to the umbra. 
% The observed spatial distribution in frequency is consistent with both emission mechanisms. 

\begin{figure}[!htb]
% \begin{mdframed}[leftmargin=5pt, rightmargin=5pt, linecolor=black!50, backgroundcolor=red!10, linewidth=2pt]
\centering
\includegraphics[width=1.\linewidth]{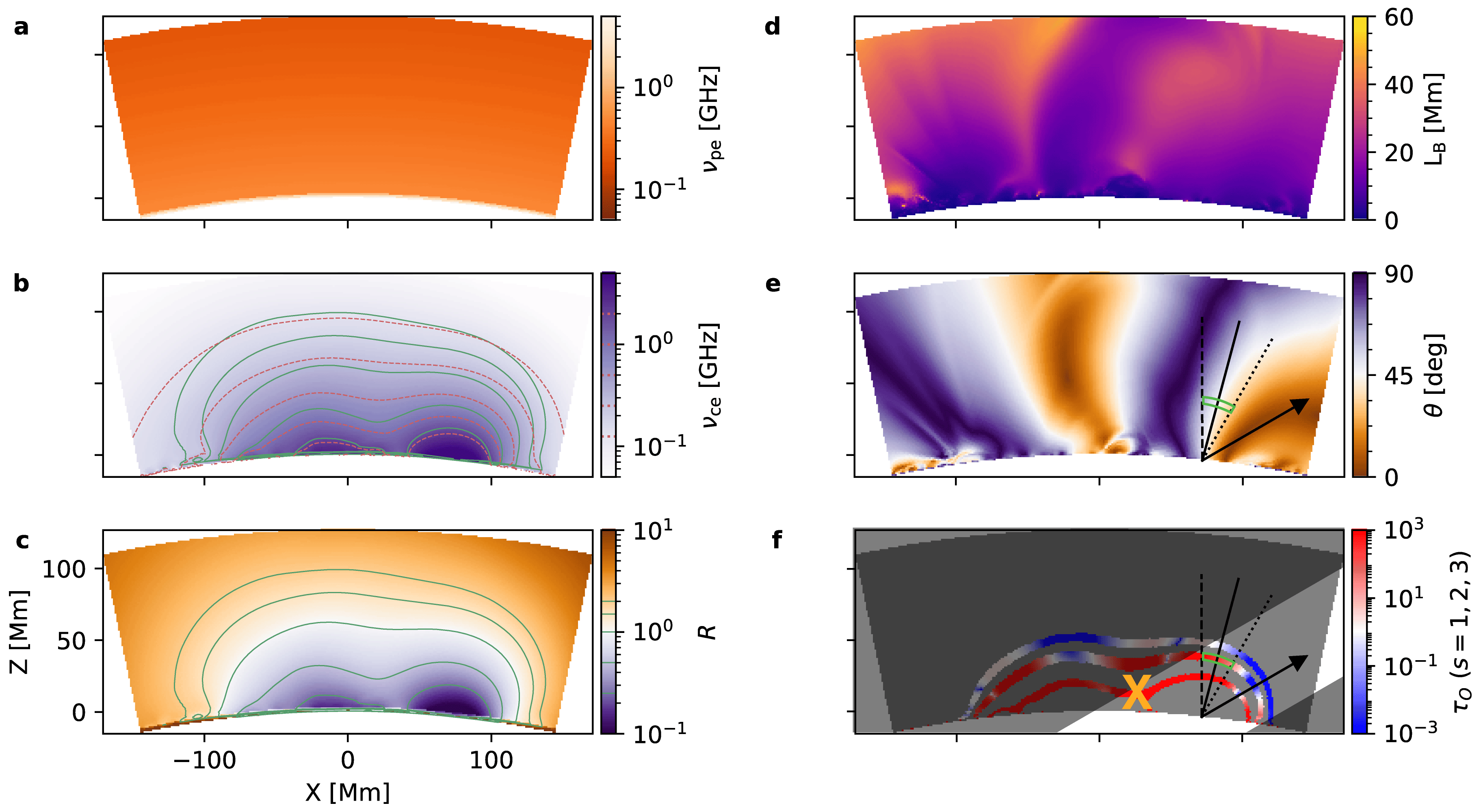}
% \end{mdframed}
\caption{\textbf{Physical characteristics in a 2D slice on the $X$--$Z$ plane of the 3D model of NOAA 12529 AR}. The location of the $X$--$Z$ plane is indicated by the semitransparent gray shade in Extended Data Fig. \ref{fig:Bmodel}(a). \textbf{a}, electron plasma frequency $\nu_\mathrm{pe}$. \textbf{b}, electron cyclotron frequency $\nu_\mathrm{ce}$. The red contours from top to bottom show the frequencies of 0.125, 0.25, 0.5, 1.0, 2.0 GHz.
% at below 1.5 GHz (0.245, 0.41, 0.61, 1.0, 1.415 GHz.) 
\textbf{c}, ratio of the electron plasma frequency to electron cyclotron frequency $R=\nu_\mathrm{pe}/\nu_\mathrm{ce}$. The green contour lines show the ratio $R$ of 0.25, 0.5, 1.0, 1.5, and 2.0, from bottom to top. The same contours are shown in panel (\textbf{b}). \textbf{d,} magnetic field scale height L$_\mathrm{B}$. \textbf{e}, magnetic field angle $\theta$ relative to the line-of-sight. \textbf{f,} gyro-resonance opacity $\tau$ of $O$ mode emission at $\nu_\mathrm{GHz}=1.0$ along the line-of-sight. The black arrow indicates the direction towards an Earth-based observer. The dotted, solid, and dashed lines indicate the direction with $\theta$ equals to $30^{\circ}$, $45^{\circ}$, and $60^{\circ}$. The corona that is transparent to the $s=3$ gyro-resonance absorption layer is outlined as the tilted white shade. The green contour denotes the $s=2$ ECM source that is inferred from the observed spatial distribution of the source locations of the coherent radio bursts. The yellow cross marks the approximate location of the solar flares.}
\label{fig:params}
\end{figure}

Although establishing a comprehensive emission model is challenging and beyond the scope of the current study due to a lack of in situ data, we can investigate the emission mechanism using remote-sensing data provided by multi-wavelength observations. By employing a data-constrained 3D coronal magnetic field model and electron number density model, we calculate the ratio $R=\nu_\mathrm{pe}/\nu_\mathrm{ce}$ in the 3D volume above the active region to identify regions more likely for the plasma radiation or ECM emission. The method has been previously employed to investigate the emission mechanism of coherent radio bursts in the solar corona\cite{2015A&A...581A...9R,2016A&A...589L...8M}. 
% We note the density model is over simplified for the highly-structured and overdense region near the flare site. 
The resulting distributions of $\nu_\mathrm{pe}$, $\nu_\mathrm{ce}$, and their ratio $R$ at the $Y=0$ plane in the model are shown, respectively, in the left column of Extended Data Fig. \ref{fig:params}. Although both the electron number density $n_e$ and magnetic field strengths $B$ decrease with increasing height in the corona, and so do the plasma frequency $\nu_\mathrm{pe}$ and cyclotron frequency $\nu_\mathrm{ce}$, the $B$ decreases more rapidly in the region above the sunspot. This leads to a low $R$ cavity just above the sunspot (Extended Data Fig. \ref{fig:params}(c)), in which the electron maser instability is likely to be excited. The ECM emission frequency $\nu$ in the $R\lesssim1$ cavity ranges from above 1.5 GHz to 250 MHz (Extended Data Fig. \ref{fig:params}(b)). This agrees with the observations that the sunspot auroral bursts are presented in a wide frequency range from $\sim$ 2 GHz all the way to the lowest frequency of RSTN at 245 MHz (Extended Data Fig. \ref{fig:supp_norp_rstn}(a)). We acknowledge that $n_e$ and $\nu_\mathrm{pe}$ are reliant on the column depth $D$, which should be regarded as a rough estimate. Nevertheless, it is important to note that the electron plasma frequency gradually varies with the fourth root of column depth $D$, following the relationship  $\nu_\mathrm{pe}\propto{n^{0.5}_e}\propto1/D^{0.25}$. Consequently, changes in $D$ have only a minimal impact on the electron plasma frequency $\nu_\mathrm{pe}$. Varying $D$ from 15 to 150\,Mm leads to a 30\% difference in $\nu_\mathrm{pe}$ with respect to the value derived using $D=50\,\mathrm{Mm}$ (Extended Data Fig. \ref{fig:NEmodel}(c)). This variation does not significantly affect the inferred ECM harmonic and the ratio $R$.

Owing to the relatively high plasma density in the solar corona compared to, e.g., planetary magnetospheres, the detectability of the ECM emission is also affected by gyro-resonant absorption due to the ambient thermal plasma at higher harmonics of the electron gyro-frequency\cite{1982ApJ...259..844M}. The opacity of the gyro-resonance layers depends on the magnetic field geometry, the thermal electron number density $n_{e}$ and the viewing direction\cite{1968SvA....12..245Z}:
\begin{equation}
\label{equ:tau}
    \tau(s, \nu, \theta)=\frac{\pi e^{2}}{2 m_{e} c} \frac{n_{e} L_{B}(\theta)}{\nu} \frac{s^{2}}{s!}\left(\frac{s^{2}}{2 \mu}\right)^{s-1} F(\theta)
\end{equation}
where $\theta$ is the angle between the LOS and the local magnetic field direction, $L_{B}(\theta)=B/|\nabla B|$ is the magnetic scale height along the LOS direction, $\mu =  m_{e} c^2/k_BT$ is the square of the ratio between the speed of light and electron thermal speed, and $F(\theta)$ is a function of angle $\theta$:
\begin{equation}
    F(\theta) = \frac{\sin^{2 s-2} \theta\left(\sin ^{2} \theta+2s \cos ^{2} \theta +\sigma\sqrt{\left. \sin ^{4} \theta+4 s^2\cos ^{2} \theta\right.}\right)^{2}}{\sin ^{4} \theta+4s^2 \cos ^{2} \theta +\sigma \sin ^{2} \theta \sqrt{ \sin ^{4} \theta+4 s^2\cos ^{2} \theta}}
\end{equation}
with $\sigma=1$ for extraordinary mode ($X$ mode) or $\sigma=-1$ for ordinary mode ($O$ mode).

\begin{figure}[!ht]
% \begin{mdframed}[leftmargin=5pt, rightmargin=5pt, linecolor=black!50, backgroundcolor=red!10, linewidth=2pt]
    \centering
    \includegraphics[width=1.0\textwidth]{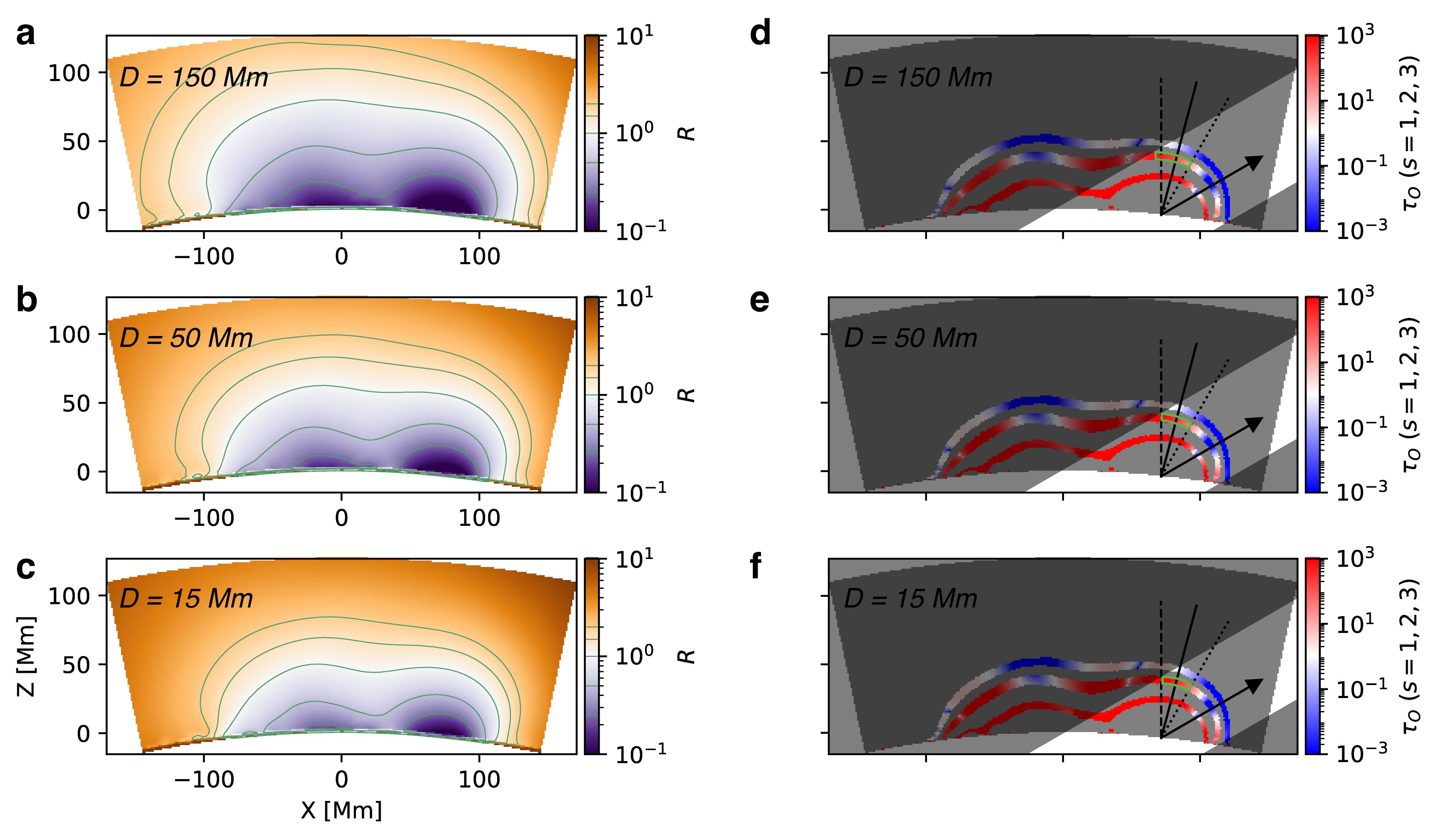}
    \caption{\textbf{Physical characteristics in a 2D slice on the $X$--$Z$ plane of the 3D model of NOAA 12529 AR for three different column depths.} The left and right columns show the same plots of the ratio $R=\nu_\mathrm{pe}/\nu_\mathrm{ce}$ and the gyro-resonance opacity $\tau$ as Extended Data Fig. \ref{fig:params}(c\&f), but for column depths $D$ of 150, 50, and 15\,Mm from top to bottom, respectively.}
    \label{fig:params_compare}
% \end{mdframed}
\end{figure}

The nearly 100\% right circularly polarized radio emission is closely associated with the sunspot in negative magnetic polarity, suggests that the emission is likely in the ordinary mode, or $O$ mode. With Equation \ref{equ:tau}, we evaluate the absorption of ECM radiation produced at a harmonic of $s_e$ by the overlying gyro-resonance layer at $s_a = s_e+1$ (the absorption at larger harmonics is much smaller and can be neglected) for a given observing frequency $\nu$. In Extended Data Fig. \ref{fig:params}(d)--(f), we show, respectively, maps of $L_{B}$, $\theta$ and the opacity $\tau$ of the $s_a=1, 2, 3$ layer for o-mode radiation at a representative frequency of $\nu_\mathrm{GHz} = 1.0$ along the line of sight at the $Y=0$ plane. Our model shows that the $s_a=2$ layer is mostly optically thick except for a narrow window at small $\theta$ values with respect to the line of sight where the ECM emission produced at the fundamental harmonic $s_e=1$ can escape. 
% This is consistent with the ECM theory\cite{1982ApJ...259..844M}. 
On the other hand, the model suggests that ECM emission produced at the second harmonic $s_e=2$ is only partially affected by the gyro-resonance absorption layer at $s_a=3$. The latter shows a morphology similar to the $s_a=2$ layer, but has a much wider transparent window covering the sunspot umbra and penumbra regions. Therefore, the second-harmonic ECM emission can escape from the transparent window in the $s_a=3$ layer and be observed, as demonstrated in the schematic picture in Fig. \ref{fig:Bmodel}(h). We note that while the opacity structure model is dependent linearly on electron number density $n_e$, the latter varies with the square root of column depth $D$. Changes in $D$ only have a minimal impact on opacity structure, following the relationship $\tau \propto D^{-0.5}$. Varying $D$ from 15 to 150\,Mm results in a scale factor of 0.7 to 2.2 for the opacity value, which is not significant enough to affect the overall opacity structure for the ECM emission and its observability (Extended Data Fig. \ref{fig:params_compare}).

We end this section with a brief discussion of the growth of ECM emission in $X$ and $O$ mode, which depends heavily on the value of $R=\nu_\mathrm{pe}/\nu_\mathrm{ce}$. Parametric investigations on ECM instability in the literature suggests that at the fundamental $O$ mode or second harmonic $X$ mode radiation would be dominant when $0.3\lesssim R\lesssim1$, and the second harmonic O-mode would dominate the radiation when $1.4\lesssim R\lesssim2$\cite{1985JGR....90.9663W,2017PhPl...24e2902T}. In the regions where the radio source at 1--2 GHz, the ratio $R$ derived from the magnetic field model and coronal density model generally falls in the range of $0.3\lesssim R\lesssim1$ (Extended Data Fig. \ref{fig:params}(b)), which would favor fundamental $O$ mode or second harmonic $X$ mode. However, as we described in the coronal density model section, the corona density depends strongly on the selection of the column depth. The absolute coronal density can be scaled up by a factor of a few if we assumed a smaller column depth, so that the ratio $R$ can satisfy required values for second harmonic $O$ mode to dominate. We caution that such uncertainty in the density model impedes an accurate determination of the ratio $R$, so as for the preferable emission wave mode. For emission below 1 GHz, we choose not to discuss the wave mode because of lacking imaging observation and polarization measurement at frequencies below 1 GHz. Nevertheless, Chen et al\cite{2013ApJ...763L..21C} found the corona may consist of many unresolved overdense loops based on the discrepancy between the density inferred from the frequency of type III radio bursts and that inferred from EUV observations. It is suggested that there are overdense magnetic loops above the sunspot with a density enhancement by up to an order of magnitude against the tenuous background, the ratio $R$ in which can be greater by the square root of the same factor. If the observed coherent radio bursts can be interpreted as ECM emission originating in these overdense loops, the second harmonic $O$ mode radiation would be more probable to dominate in these loops with a higher plasma density. The overdense loops scenario offers an alternative avenue for second harmonic $O$ mode ECM radiation to dominate in a general low $R$ environment ($R$<1).
% , alleviating the discrepancy between the model and the ECM theory.

\begin{figure}[!htb]
% \begin{mdframed}[leftmargin=5pt, rightmargin=5pt, linecolor=black!50, backgroundcolor=red!10, linewidth=2pt]
\centering
\includegraphics[width=0.7\linewidth]{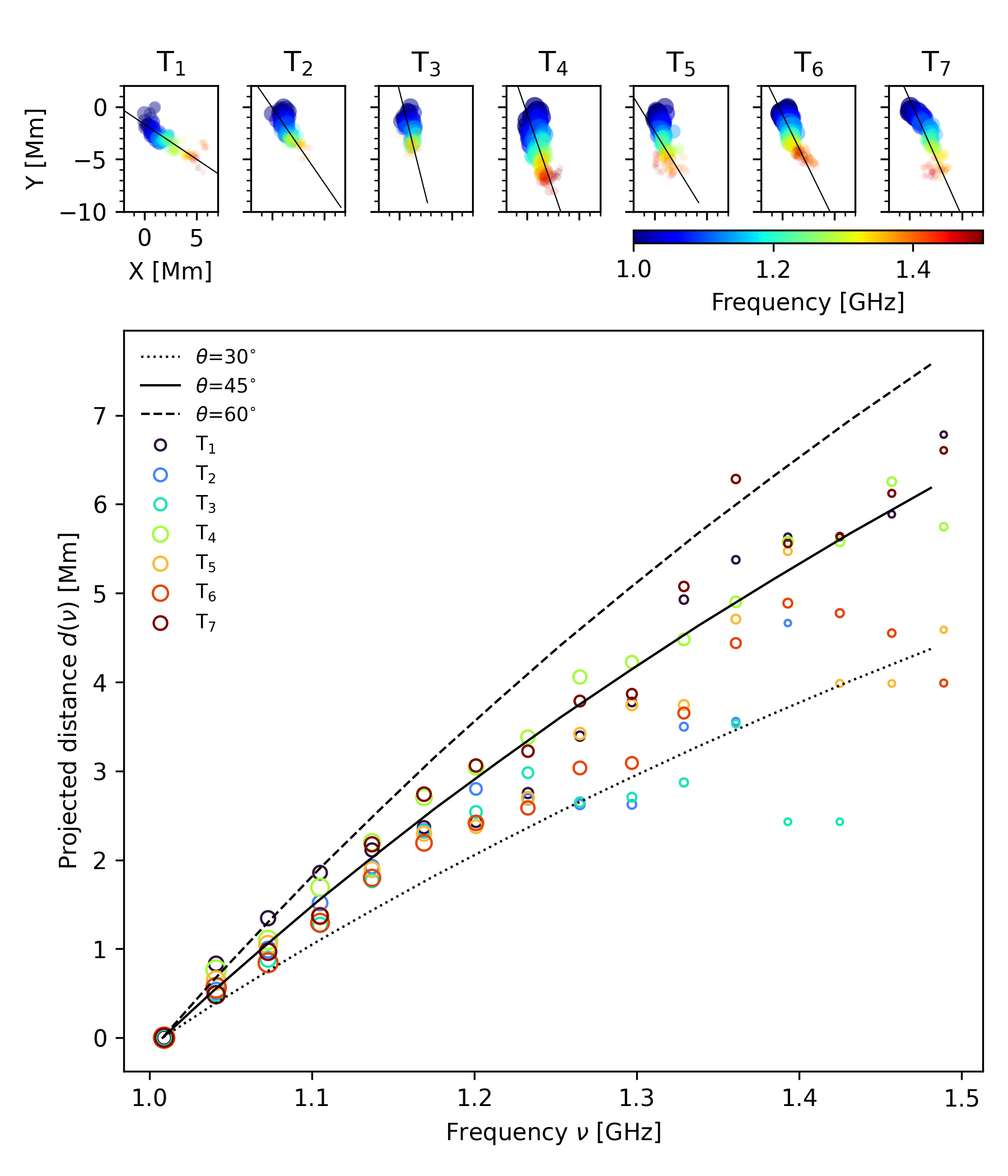}
% \end{mdframed}
\caption{\textbf{Distribution of the frequency-dependent source centroid locations for seven individual 20-second time integrations.} \textbf{Top panel}: The frequency distribution of radio source locations for the example times (same as these used for Fig. \ref{fig:Bmodel}a). The close-to-linear distribution of radio sources in space as a function of frequency is attributed to radio sources along the respective loop, which is outlined by a solid line. \textbf{Bottom panel}: the plane-of-the-sky projected distance $d(\nu)$ of radio source locations from the source centroid at 1 GHz along the direction of the flux tubes. The predicted $d(\nu)$ dependence along flux tubes of three orientations with viewing angle $\theta$ equals to $30^{\circ}$, $45^{\circ}$, and $60^{\circ}$ as the dotted, solid, and dashed lines, respectively.}
\label{fig:dis_freq}
\end{figure}

\subsection*{Spatial distribution of the ECM source}
With spatially and temporally resolved imaging spectroscopy, we can derive the centroid location of each observed source at any given time $t$ and frequency $\nu$, provided that it has a sufficient signal-to-noise-ratio against the quiescent background. The top panels in Extended Data Fig. \ref{fig:dis_freq} shows examples of the distribution of the frequency-dependent source centroid locations for seven individual 20-second time integrations. Intriguingly, the spatial distribution of each burst forms a close-to-linear distribution in a way similar to those derived from electron-beam-driven type III radio bursts \cite{2013ApJ...763L..21C,2018ApJ...866...62C}.

As the emission frequency $\nu$ due to the ECM mechanism depends only on the magnetic field strength $B$, the observed spatial distribution of the radio source centroid location at a given frequency $\nu$ in projection $\vec{r}(\nu)$ is governed by the magnetic field distribution above the sunspot $\vec{r}(B)$. Our observations have suggested that, for each radio burst, the distribution of the frequency-dependent radio source centroids is likely aligned with a single magnetic loop. Therefore, we can reduce the distribution of the radio sources to a one-dimensional dependence of $d(\nu)$, where $d$ denotes the plane-of-the-sky projected distance from a reference location along the respective loop, selected as that of the source centroid at 1 GHz. We construct the $d(\nu)$ dependence of the seven example times, shown in the bottom panel of Extended Data Fig. \ref{fig:dis_freq} as colored circles (whose sizes are scaled by their peak flux). With the 3D coronal magnetic field model to provide the mapping of $d(B)$ for each magnetic loop in projection, we can predict the expected $d(\nu)$ dependence in different viewing angles of $\theta$. In the bottom panel in Extended Data Fig. \ref{fig:dis_freq}, we show the predicted $d(\nu)$ dependence along three directions with $\theta$ equals to $30^{\circ}$, $45^{\circ}$, and $60^{\circ}$ as the dotted, solid, and dashed lines, respectively. The results suggest that the viewing angle $\theta$ of the radio sources with the observed $d(\nu)$ dependence should lie in the range of $30^{\circ}$--$60^{\circ}$. This range of viewing angles demarcates a region in the vicinity of the apex of the $s=2$ gyroresonance dome (denoted as the green contour in Extended Data Fig. \ref{fig:params}(e\&f)), corresponding to the predicted ECM emission sites that are co-spatial with the transparent window of the $s_a=3$ gyroresonance layer in projection. The range of viewing angles is also consistent with the values of $\theta$ in the inferred source region (green contour in Extended Data Fig. \ref{fig:params}(e).)
% Therefore, we argue this excellent agreement in the viewing angle derived using two independent approaches is a strong support for the ECM interpretation of the observed radio bursts. 
We note that the beaming angle of 30$^{\circ}$--60$^{\circ}$ is generally smaller than observed values (75$^{\circ}$--80$^{\circ}$) in the case of Jovian radio auroral radiation\cite{2008JGRA..113.3209H}, but similar to the beaming angles ($\gtrsim 40^{\circ}$) of the auroral kilometric radiation in the terrestrial case\cite{2000ApJ...538..456E}. The latter is often attribute to the refraction effect that the radio waves is initially emitted to the perpendicular direction but is ducted away through refraction at the borders of the density depletion cavity in which the emission is produced\cite{2000ApJ...538..456E}. This does not seem to be the case for the sunspot radio emission. For the ducting to occur, the plasma frequency of the surrounding boundary of the low-density cavity should be significantly higher than the emitting frequency of the radiation. This requires a plasma density of at least a few times of $10^{10}\,\mathrm{cm}^{-3}$, which is much higher than the inferred plasma density near the source region. One possibility is that the small beaming angle is a result of ECM processes coupled with Alfv\'en waves. Wu et al (2012)\cite{2012PhPl...19h2902W} found that preexisting Alfv\'en waves can modify the classical ECM processes by affecting the velocity distribution of energetic electrons via pitch-angle scattering.
% For second harmonic radiation, beaming angles that smaller than 50$^{\circ}$ can be achieved.
% under the condition of a loss-cone distribution plus a electron beam (also known as ``ring-beam'' distribution)\cite{2017PhPl...24e2902T}. The ring-beam distribution is not an infrequent feature for energetic electrons accelerated in astrophysical plasma. For instance, fast electron beams associated with solar type-III bursts \cite{2013ApJ...763L..21C,2018ApJ...866...62C} may have a ring component if they are injected to the magnetic field line with an initial pitch angle.  
Indeed, Alfv\'en waves is expected in the quiet solar corona\cite{2011Natur.475..477M, 2018NatPh..14..480G} and flare regions\cite{2019ApJ...872...71Y}. The beaming angle of second harmonic $O$ mode ECM emission, if driven by nonthermal electrons in a horseshoe distribution, can be as low as 30$^\circ$. The interpretation of the observed coherent radio bursts as second harmonic $O$ mode ECM emission is consistent with the small beam opening angles predicted in the analyticla study\cite{2017PhPl...24e2902T}. Nevertheless, we acknowledge that the observed beaming angle is not fully consistent with the values of previously observed planetary ECM emissions. To produce a fully self-consistent model of ECM emission from the sunspot requires not only a better understanding the corona context using a data constrained MHD model, but also a fully kinetic simulations to achieve a realistic nonthermal electron distribution, both of which are beyond the scope of this study.
% The typical spatial dispersion in frequency is about $\sim 12\,\mathrm{Mm/GHz}$.
% % , or $\sim16^{\prime\prime}/\mathrm{GHz}$. 
% This value is more than one order of magnitude lower than the one reported in the type IIIdm paper (Chen et al 2013). This corresponds to a density scale height of 4 Mm, or a coronal temperature of 0.08 MK, assuming plasma radiation and hydrostatic equilibrium.

On the other hand, plasma emission radiates at a frequency close to the fundamental or second harmonic electron plasma frequency $\nu_\mathrm{pe}$. Plasma radiation at frequencies in the range 1-1.5 GHz suggests the ambient plasma density above the sunspot in the range $1$--$3\times10^{10}\,\mathrm{cm^{-3}}$, which is more than an order of magnitude higher than coronal density model constrained by the DEM analysis (Extended Data Fig. \ref{fig:NEmodel}(c)). It is worth asking, however, whether the observed radio bursts are plasma radiation originating from the aforementioned overdense magnetic loops in the corona. As the emission frequency $\nu$ due to the plasma radiation mechanism is proportional to the square root of plasma density ($\nu_\mathrm{pe}\propto \sqrt{n_\mathrm{e}}$), the $d(\nu)$ of the plasma radiation source is determined by the plasma density distribution above the sunspot. The plasma frequency $\nu$ varies from 1.0 to 1.5 GHz over $d(\nu)$ of 3--6 Mm corresponds to the plasma density varying from $1\times10^{10}$ to $3\times10^{10}\,\mathrm{cm^{-3}}$ over a height of 4--8 Mm, assuming the overdense loops have inclination angles of $\sim45^\circ$. This is equivalent to a density scale height $\mathrm{L}_{n}=n_\mathrm{e}/(dn_\mathrm{e}/d\vec{r})$ of 4--8 Mm under hydrostatic equilibrium, which are at least an order of magnitude smaller than the hydrostatic value at typical coronal temperature ($\mathrm{L}_{n}\approx46T_\mathrm{MK}\,\mathrm{Mm}$, or 94 Mm for $T = 2\,\mathrm{MK}$\cite{2005psci.book.....A}). Such steep density profiles, in other words, steep pressure gradients, are inconsistent with the expectation for the quasi-static magnetic structures above the sunspot. Therefore, we reaffirm the conclusion that plasma radiation is unlikely the responsible mechanism for the observed radio bursts.

\subsection*{Observability of the sunspot ECM emission}
ECM generates radiation along a thin, conical sheet. Consequently, sunspot ECM emissions are only observable when the narrow radiation beam crosses the line of sight due to the Sun's rotation. Although ECM emissions should also appear when the sunspot is in the symmetric phase on the western side of the solar disk, they are only visible for a specific heliocentric angle range on the eastern side, as shown in Extended Data Fig. \ref{fig:supp_radio_profile}. In this study, we attribute this phenomenon to the influence of solar flares on the observability of sunspot ECM emissions. During the observed sunspot radio bursts, the emission source region was far from the flare sites. However, as the sunspot rotates to the other side of the solar disk, the potential emission source (the green contour line in Extended Data Fig. \ref{fig:params_varyingtheta}) would become much closer to the flare sites. The close spatial proximity of source sites to flares can significantly alter the local plasma and magnetic environment, affecting ECM generation and escape. A more detailed discussion of various impacts is provided below:

\begin{figure}[!htb]
% \begin{mdframed}[leftmargin=5pt, rightmargin=5pt, linecolor=black!50, backgroundcolor=red!10, linewidth=2pt]
\centering
\includegraphics[width=0.5\linewidth]{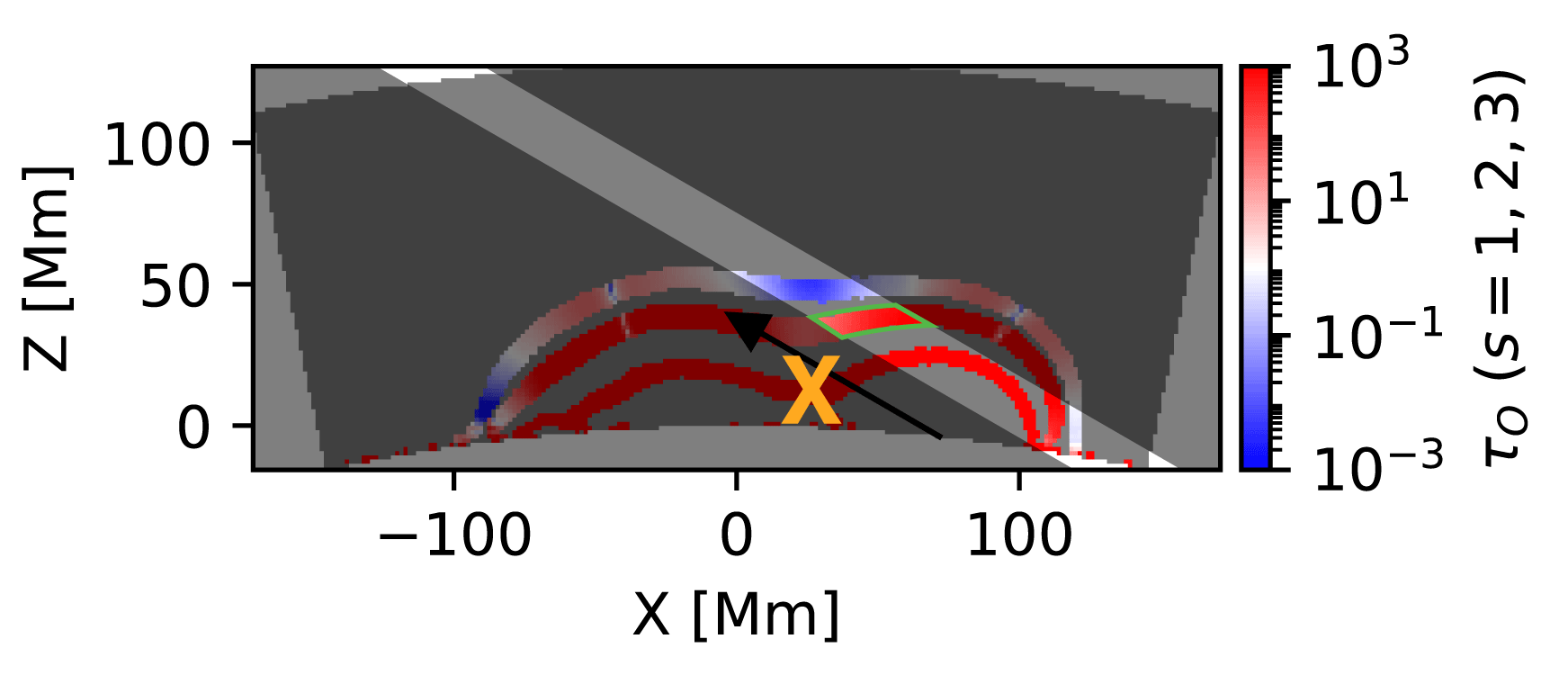}
% \end{mdframed}
\caption{Similar as Extended Fig. \ref{fig:params}(f), this figure displays the gyro-resonance opacity $\tau$ of $O$ mode emission at $\nu_\mathrm{GHz}=1.0$ along the line-of-sight, considering the AR is located to the opposite longitude with respect to the solar disk center and assuming a 10-fold increase in electron number density $n_\mathrm{e}$ to account for the impact of the nearby flare region. The black arrow indicates the direction towards an Earth-based observer. The corona that is transparent to the $s=3$ gyro-resonance absorption layer is outlined as the tilted white shade. The green contour represents the potential source location of the $s=2$ ECM radio emissions. The yellow cross marks the approximate location of the solar flares.}
\label{fig:params_varyingtheta}
\end{figure}

\begin{itemize}
    \item ECM condition: The ECM mechanism requires a specific condition of $R\lesssim1$ at the source site \cite{1982ApJ...259..844M}. Given the source site's proximity to flares, the impact of solar flares on coronal density structures must be considered. The thermal plasma density distribution during flares is highly dynamic and complicated, which required specialized modeling efforts to simulate its temporal evolution\cite{2019NatAs...3..160C}. While this is beyond the scope of this study, a rough estimate can still be performed. Solar flares can increase coronal plasma density by a factor of over 10 within tens of seconds\cite{2007ApJ...668.1210R}. This over 10-fold increase in density near flare regions could raise the $R$ value in the nearby ECM source region by a factor of a few, as $R$ is proportional to $\sqrt{n_\mathrm{e}}/B$. In this case, the source region near the flare sites may not meet the ECM condition and cannot produce ECM emission.
    
    \item Opacity: The solar corona's highly structured nature means that the opacity of the gyro-resonance absorption layer, which depends on the viewing direction and 3D distributions of the coronal magnetic field and thermal plasma\cite{1982ApJ...259..844M}, will change over time as the Sun rotates. This change subsequently affects the emissivity and polarization of sunspot ECM emissions. As mentioned earlier, the opacity varies linearly with the electron number density $n_\mathrm{e}$. A more than 10-fold increase in $n_\mathrm{e}$ near the flare regions could significantly alter the opacity of the nearby gyro-resonance layer in the line of sight (Extended Data Fig. \ref{fig:params_varyingtheta}), restricting the escape of ECM radiation.
    
    \item Magnetic configuration: In our interpretation, the generation of the sunspot radio bursts requires closed magnetic loops trapping energetic electrons. Recurring non-eruptive flares are favored because they can continuously replenish the population of energetic electrons in the overarching radio-hosting magnetic field lines above the energy release site without disrupting the overall magnetic field geometry. In contrast, eruptive flares are expected to open up the magnetic field lines above \cite{2015SoPh..290.3457S}. The newly accelerated electrons by solar flares can easily escape to interplanetary space along the opened field lines without getting trapped in the low corona. Moreover, the opening of the overarching magnetic field lines results in the destruction of the electron reservoir where energetic electrons accelerated by previous flares are trapped, further interrupting the generation of the sunspot radio bursts. As demonstrated in Extended Data Fig. \ref{fig:supp_radio_profile}(e), NOAA 12529 AR became more violent on 2016 April 14, producing an M class flare and a few sub-M class flares. In contrast, the AR only produced a few C class flares earlier.
    
\end{itemize}

One or more of the factors mentioned above may influence the observability of the sunspot ECM emission as the viewing angle varies, in addition to the effect of stellar rotation. Understanding the viewing geometry effects requires long-term broadband dynamic imaging spectropolarimetry to study in detail the time-dependent variations of the morphology, location, and polarization of the radio source. The emissivity of sunspot coherent radiation, including but not limited to the emission intensity, polarization and wave mode, depends heavily on the magnetic field distribution and coronal plasma distribution, which are subjected to change with time due to the gradual evolution of the radio-hosting active region and abrupt energy release events. To establish the relationships of coherent radio bursts with the structure and dynamics of the corona, including plasma density inhomogeneity, magnetic field topologies, and their variations, broadband dynamic imaging spectroscopy observations in the frequency range of 200 MHz to 2 GHz are required. These observations should be complemented by other multi-wavelength observations, including optical, EUV, and X-ray. Moreover, with the aid of data-constrained coronal 3D magnetic and thermodynamic structure, imaging spectropolarimetry would allow us to characterize and constrain coronal thermal parameters, magnetic field properties, and the energy distribution of nonthermal electrons that drive ECM emission. Detailed context data at other wavelengths, coupled with radio imaging observations, would enable the examination of the relationship between the rotational modulations of sunspot ECM emission and other significant observables, such as photometric and Doppler spectroscopic measurements\cite{2020ApJ...897..125R, 2022ApJ...925..125D}. Such observations will be crucial for improving our understanding of the generation and rotational modulation effects of ECM bursts from the Sun and other stars.

An important corollary to the discussion of observability of the sunspot ECM emission is the potential influence of viewing effects on the duty cycle of stellar ECM emissions. If these stellar emissions were associated with the presence of starspots on the surfaces of these (sub)stellar systems, viewing effects may play a role in determining the temporal pattern within a single duty cycle of these emissions. Like the solar case, many previously reported stellar ECM emissions only appear in a certain range of stellar rotational phase\cite{2009ApJ...695..310B,2022ApJ...925..125D}. Depending on the relative positions of the starspot and its neighboring flare sites with respect to the light of sight, as well as the inclination of the rotation axis relative to the line of sight as viewed from the Earth, stellar ECM emissions may manifest a second time in symmetric phase with respect to the stellar-centric $0^{\circ}$ longitude line. Similar cases have also been observed in other stars\cite{2011ApJ...739L..10T,2019MNRAS.488..559Z}. Furthermore, as a type of transient radio source, the duty cycle of stellar ECM emissions could also vary as a consequence of the short-term evolution of the starspot and its magnetic connectivity to other parts of the stellar surface. Long-term variations, such as  the emergence of a starspot, may lead to the appearance of ECM radio emission on a star that previously lacked such radio bursts in earlier observations, and vice versa. Understanding these viewing effects and their impact on the observability of sunspot and starspot ECM emissions can inform future survey-type observations of stellar ECM emissions.

In concluding this section, we explore the potential applicability of our model to early-type stars (spectral type B or A). Rotationally modulated ECME emissions have been discovered in several early-type stars, exhibiting consistent phases over extended periods \cite{2011ApJ...739L..10T,2022ApJ...925..125D}. The prevailing hypothesis suggests that global dipole-like magnetic fields can create magnetic-mirror-like conditions necessary for maser emission\cite{2004A&A...418..593T}. These stars often possess magnetic fields with kG strengths and simple topologies, often well approximated by a dipole\cite{2019A&A...621A..47K}. Stellar wind\cite{2011ApJ...739L..10T} or magnetic reconnection in the magnetosphere\cite{2020MNRAS.493.4657L} may provide energetic electrons required to drive ECM emissions. These electrons then become trapped in the stellar magnetospheric pole regions, producing magnetic-mirror-like conditions that facilitate ECME emission. However, recent studies reveal that radio ECM emission can emerge from stars with surface magnetic fields significantly deviating from an axisymmetric dipole\cite{2014A&A...565A..83K}, leaving the role of the dipole magnetic field topology in producing radio ECM emission remains an open question. Analogous to the magnetospheric pole regions and the electromagnetic engine in the magnetosphere, a localized magnetic structure such as a starspot and nearby flare activity can also produce radio ECM emissions, similar to the solar case. Recent time-series photometry from space missions like Kepler, K2, and TESS challenges this belief, suggesting that rotational modulation is probable in stars with radiative envelopes \cite{2019MNRAS.485.3457B}. The observed light variations can be most simply interpreted as rotational modulation induced by surface brightness inhomogeneities, such as magnetic spots. These localized magnetic fields might be associated with magnetic fields generated in subsurface convection zones \cite{2011A&A...534A.140C}. Additionally, it is plausible that differential rotation in A and B stars may be sufficient to create local magnetic fields via dynamo action \cite{2004A&A...422..225M}. Due to the extreme stability of the magnetic fields in these types of stars\cite{2018MNRAS.475.5144S}, the periodic ECME emission can be remarkably stable, unlike the case for the highly dynamic magnetic field of the Sun. If the rotational modulation observed in early-type stars indeed results from persistent magnetic spots, our model could offer an alternative approach to generating a stable burst duty cycle over extended periods in these stars. However, this proposition depends on the presence of persistent magnetic spots on early-type stars. Further investigation is necessary to assess the efficacy of the model in accounting for the long-term radio emission behavior exhibited by these stars.

\subsection*{Source size and brightness temperature}
The broadband radio bursts emitting from the sunspot are detectable across frequencies ranging from 245 MHz to 1.7 GHz. The maximum flux density of these recorded bursts in stokes \textit{I}, observed at 245 MHz at approximately 23 UT on April 12, is around 2000 sfu, or 2$\times10^{10}$ mJy. Were the observed bursts to be viewed from a stellar distance, the flux density at 245 MHz can be expressed as $0.5 (D_{\rm pc})^{-2}$ mJy, where $D_{pc}$ is the stellar distance in parsec. If we take the median distance of 20 parsec in the recent LOFAR observations\cite{2021NatAs...5.1233C}, the flux density would approximate 1 $\mu$Jy. Although the total flux density is two or three orders of magnitude lower than the median values reported in the recent LOFAR observations at hundreds of MHz \cite{2021NatAs...5.1233C}, 
\BC{we argue that the brightness temperature of the sunspot radio source can still be very high. 
%(For the M dwarf bursts, if assuming the source (at any given frequency) is the size of the entire stellar disc, the brightness temperature can reach $10^{14}$ K.) 
This is owing to the likely compact and unresolved nature of many individual bursts that comprise the observed radio source.} 
We consider the magnetic loops anchored at the sunspot where impulsively accelerated electrons are injected and an unstable loss-cone distribution is set up by magnetic mirroring on the sunspot end of the loops. the brightness temperature for a continuously operating maser is given by\cite{1982ApJ...259..844M}, 
\begin{equation}
    T_{\mathrm{b}}=\frac{m_e v_0^2}{4 \pi k_{\mathrm{B}}} \frac{c^2}{\nu^2 L r_0}  \approx 2 \times 10^{12}\left(\frac{E}{1 \mathrm{~keV}}\right)\left(\frac{\nu}{200 \mathrm{~MHz}}\right)^{-2}\left(\frac{L}{R_\odot}\right)^{-1} \mathrm{~K}
\end{equation}
where $v_0$ is the velocity of the emitting electrons, $\nu$ is the emission frequency, $L$ is the length scale of the magnetic trap with respect to the solar radius $R_\odot$, $r_0$ is the classical electron radius, $m_e$ is the electron mass, $k_{\mathrm{B}}$ is the Boltzmann constant, and $c$ is the speed of light.  Using a magnetic loop length scale of approximately 1/10 of the solar radius $R_\odot$ and electron energy $E$ of 20 keV, the brightness temperature is estimated to be approximately $10^{14}$ K at 245\,MHz and $10^{13}$ K at 1\,GHz for the fundamental emission. For the second harmonic ECM emission, the values are $10^{12}$ K at 245 MHz and $10^{11}$ K at 1 GHz, considering the latter's growth rate potentially being two orders of magnitude lower than the fundamental\cite{2017PhPl...24e2902T}. 
% The inferred brightness temperature for the second harmonic ECM emission aligns with the recently reported ECM bursts from M dwarfs \cite{2020NatAs...4..577V,2021NatAs...5.1233C}.
% \SYS{The inferred brightness temperature for the second harmonic ECM emission aligns with the observed brightness temperature .}

\begin{figure}[!htb]
% \begin{mdframed}[leftmargin=5pt, rightmargin=5pt, linecolor=black!50, backgroundcolor=red!10, linewidth=2pt]
\centering
\includegraphics[width=\linewidth]{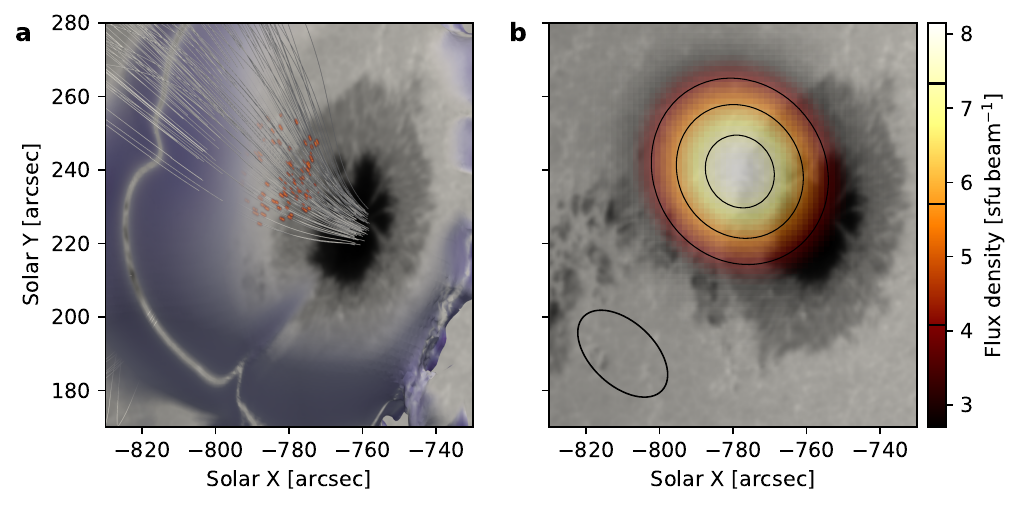}
% \end{mdframed}
\caption{\textbf{Simulation of the 20\,sfu sunspot radio image at 1 GHz.} \textbf{a}: The input model of the radio ECM sources at 1 GHz displayed in a context of the sunspot and the corresponding gyroresonance opacity layer, similar to Fig. \ref{fig:Bmodel}h. The model consists of a hundred randomly distributed point sources with a brightness temperature of $10^{11}$ K at selected field lines. \textbf{b}: similar Extended Data Fig. \ref{fig:supp1}, but showing the simulated VLA image at $\nu_\mathrm{GHz} =1.0$ overlaid on the HMI continuum intensity image.} 
\label{fig:tb}
\end{figure}

High brightness temperatures are not uncommon for solar radio bursts, as evidenced by solar decimetric spikes\cite{1986SoPh..104...99B} whose source size scale is suggested to be below 1000 km. For the 20 sfu radio observed by VLA at 1 GHz, the brightness temperature corresponds to a source size of 1000$\times$1000 km. Since the observed radio source of the 1\,GHz image is marginally resolved by VLA, a single 1000$\times$1000 km source can not account for the asymmetric structure seen in the VLA 1 GHz image. We speculate that the observed radio source may comprise many small, unresolved sources instead of a single 1000$\times$1000 km source. To illustrate such a possibility, we employed a hundred ($N=100$) randomly distributed $10^{11}$ K point sources in the second harmonic gyroresonance of $\nu_\mathrm{GHz}=1$ at selected field lines. We note that the choice of the source number $N$ is arbitrary, but it does not affect our subsequent analysis. Each point source has a uniform diameter of $1000/\sqrt{N}\mathrm{~km}$ and a flux density of $20/N$ sfu. The model is shown in Extended Data Fig. \ref{fig:tb}a. The input model is first convolved with a $40''$ Gaussian beam to account for angular scattering at 1 GHz by coronal turbulence\cite{1994ApJ...426..774B}. Subsequently, the model is convolved with the synthesis beam of the VLA at 22 UT on April 9 to create a simulated VLA image. The result, shown in Extended Data Fig. \ref{fig:tb}b, qualitatively agrees with the actual VLA image at this frequency in terms of source location and size. It is important to note that the argument for multiple sources is consistent with the inhomogeneous and fragmentary nature of magnetic reconnection, as evidenced by recent VLA solar observations\cite{2018ApJ...866...62C}. Under this scenario, semi-relativistic electron beams can be emanated into distinct field lines, even though they are accelerated within an extremely compact region ($\sim600 \mathrm{~km}$) in the low solar corona over a brief time scale (1--2 seconds). Assuming that each ECM source is associated with a different electron beam energized within the same acceleration region, they can, in principle, arrive at various field lines above the sunspot, resulting in the scattered distribution of ECM sources as portrayed in the model. \SYS{While our observations show that the radio bursts could originate from numerous distinct magnetic field lines, we note that these field lines are all rooted at the umbra region of the sunspot with a negative polarity. Consequently, at any particular instance, these field lines share similar line-of-sight angles with regard to the observer, which gives rise to a high degree of circular polarization of the bursts over our observational period of several hours.} %are virtually uniform. Provided that the magnetic polarity of the sunspot remains constant, the detected polarization remains unaffected. In circumstances where radio ECM emissions emanate from tenacious starspots over prolonged durations---potentially spanning several years or even decades\cite{2020ApJ...897..125R}---it is anticipated that the radio emission will sustain the same sense of circular polarization over an extended observation period\cite{2019ApJ...871..214V}.}

% \BC{[BC: The following discussions are good but, as the referee suggested, they deviate from the main point of our article and create many ``holes'' that we do not necessarily want to patch. I suggest replacing them with a brief paragraph, something like below.]}

\BC{Nevertheless, explaining the observed radio bursts on certain M dwarfs with orders of magnitude higher flux density than our case remains challenging. However, we note that the plasma environment on these stars are much more extreme than that on the Sun. First, the average surface magnetic field on these stars can exceed kilo Gauss \cite{1996ApJ...459L..95J,2010MNRAS.407.2269M,2021A&ARv..29....1K}, compared to $\sim$1--10 Gauss on the Sun. The magnetic fields in these starspots may exceed those in sunspots by orders of magnitude, potentially boosting the intrinsic brightness of the ECM emission\cite{2018ApJS..237...25K}. Second, the level of flare activity on these fully convective stars is much higher than that on the Sun, allowing a possibility to pertain a much larger spatial and temporal filling factor for these bursts over the stellar disk. This speculation is reflected in the radio/X-ray correlation in the context of the GB relation, where the sunspot radio source exhibits similar deviation from the radio/X-ray relation as other stellar radio ECM emissions, despite its relatively low magnetic activity level (X-ray luminosity) among the others. Both may potentially contribute to the much brighter bursts observed on these active stars, although further observational, theoretical, and modeling investigations are required.}

\section*{Data availability}
The data that support the plots and other findings within this paper are available at \url{https://data.nrao.edu/portal} (VLA; Project ID: 16A-377), \url{https://www.sws.bom.gov.au/World_Data_Centre} (RSTN), \url{https://solar.nro.nao.ac.jp/norp} (NoRP), \url{https://www.swpc.noaa.gov/products/goes-x-ray-flux} (GOES), and  \url{http://jsoc.stanford.edu} (SDO), or from the corresponding author upon reasonable request.

\section*{Code availability}
The magnetic field extrapolation\cite{2004SoPh..219...87W} software packages are available through IDL SolarSoft at \url{https://sohowww.nascom.nasa.gov/solarsoft}. The regularized inversion code for differential emission measure (DEM) calculation\cite{2012A&A...539A.146H} is available at \url{https://github.com/ianan/demreg}. Public software packages utilized in this study include SunCASA \url{https://github.com/suncasa/suncasa-src}, CASA\cite{2007ASPC..376..127M} \url{https://casa.nrao.edu}, SunPy\cite{2020ApJ...890...68S} \url{https://sunpy.org}, and Astropy\cite{2022ApJ...935..167A} \url{https://www.astropy.org}.

%TC:endignore

%\bibliography{reference}

\begin{thebibliography}{10}
\urlstyle{rm}
\expandafter\ifx\csname url\endcsname\relax
  \def\url#1{\texttt{#1}}\fi
\expandafter\ifx\csname urlprefix\endcsname\relax\def\urlprefix{URL }\fi
\expandafter\ifx\csname doiprefix\endcsname\relax\def\doiprefix{DOI: }\fi
\providecommand{\bibinfo}[2]{#2}
\providecommand{\eprint}[2][]{\url{#2}}

\bibitem{1998JGR...10320159Z}
\bibinfo{author}{{Zarka}, P.}
\newblock \bibinfo{journal}{\bibinfo{title}{{Auroral radio emissions at the
  outer planets: Observations and theories}}}.
\newblock {\emph{\JournalTitle{\jgr}}} \textbf{\bibinfo{volume}{103}},
  \bibinfo{pages}{20159--20194} (\bibinfo{year}{1998}).

\bibitem{2005ApJ...627..960B}
\bibinfo{author}{{Berger}, E.} \emph{et~al.}
\newblock \bibinfo{journal}{\bibinfo{title}{{The Magnetic Properties of an L
  Dwarf Derived from Simultaneous Radio, X-Ray, and H{\ensuremath{\alpha}}
  Observations}}}.
\newblock {\emph{\JournalTitle{\apj}}} \textbf{\bibinfo{volume}{627}},
  \bibinfo{pages}{960--973} (\bibinfo{year}{2005}).

\bibitem{2007ApJ...663L..25H}
\bibinfo{author}{{Hallinan}, G.} \emph{et~al.}
\newblock \bibinfo{journal}{\bibinfo{title}{{Periodic Bursts of Coherent Radio
  Emission from an Ultracool Dwarf}}}.
\newblock {\emph{\JournalTitle{\apjl}}} \textbf{\bibinfo{volume}{663}},
  \bibinfo{pages}{L25--L28} (\bibinfo{year}{2007}).

\bibitem{2022ApJ...935...99B}
\bibinfo{author}{{Bastian}, T.~S.}, \bibinfo{author}{{Cotton}, W.~D.} \&
  \bibinfo{author}{{Hallinan}, G.}
\newblock \bibinfo{journal}{\bibinfo{title}{{Radio Emission from UV Cet:
  Auroral Emission from a Stellar Magnetosphere}}}.
\newblock {\emph{\JournalTitle{\apj}}} \textbf{\bibinfo{volume}{935}},
  \bibinfo{pages}{99} (\bibinfo{year}{2022}).

\bibitem{2015Natur.523..568H}
\bibinfo{author}{{Hallinan}, G.} \emph{et~al.}
\newblock \bibinfo{journal}{\bibinfo{title}{{Magnetospherically driven optical
  and radio aurorae at the end of the stellar main sequence}}}.
\newblock {\emph{\JournalTitle{\nat}}} \textbf{\bibinfo{volume}{523}},
  \bibinfo{pages}{568--571} (\bibinfo{year}{2015}).

\bibitem{2019MNRAS.488..559Z}
\bibinfo{author}{{Zic}, A.} \emph{et~al.}
\newblock \bibinfo{journal}{\bibinfo{title}{{ASKAP detection of periodic and
  elliptically polarized radio pulses from UV Ceti}}}.
\newblock {\emph{\JournalTitle{\mnras}}} \textbf{\bibinfo{volume}{488}},
  \bibinfo{pages}{559--571} (\bibinfo{year}{2019}).

\bibitem{2019BAAS...51c.484K}
\bibinfo{author}{{Kao}, M.} \emph{et~al.}
\newblock \bibinfo{journal}{\bibinfo{title}{{Magnetism in the Brown Dwarf
  Regime}}}.
\newblock {\emph{\JournalTitle{\baas}}} \textbf{\bibinfo{volume}{51}},
  \bibinfo{pages}{484} (\bibinfo{year}{2019}).

\bibitem{1998A&A...331..596B}
\bibinfo{author}{{Benz}, A.~O.}, \bibinfo{author}{{Conway}, J.} \&
  \bibinfo{author}{{Gudel}, M.}
\newblock \bibinfo{journal}{\bibinfo{title}{{First VLBI images of a
  main-sequence star}}}.
\newblock {\emph{\JournalTitle{\aap}}} \textbf{\bibinfo{volume}{331}},
  \bibinfo{pages}{596--600} (\bibinfo{year}{1998}).

\bibitem{2023arXiv230212841K}
\bibinfo{author}{{Kao}, M.~M.}, \bibinfo{author}{{Mioduszewski}, A.~J.},
  \bibinfo{author}{{Villadsen}, J.} \& \bibinfo{author}{{Shkolnik}, E.~L.}
\newblock \bibinfo{journal}{\bibinfo{title}{{Resolved imaging of an extrasolar
  radiation belt around an ultracool dwarf}}}.
\newblock {\emph{\JournalTitle{\nat}}}  (\bibinfo{year}{2023}).

\bibitem{2009ApJ...695..310B}
\bibinfo{author}{{Berger}, E.} \emph{et~al.}
\newblock \bibinfo{journal}{\bibinfo{title}{{Periodic Radio and
  H{\ensuremath{\alpha}} Emission from the L Dwarf Binary 2MASSW
  J0746425+200032: Exploring the Magnetic Field Topology and Radius Of An L
  Dwarf}}}.
\newblock {\emph{\JournalTitle{\apj}}} \textbf{\bibinfo{volume}{695}},
  \bibinfo{pages}{310--316} (\bibinfo{year}{2009}).

\bibitem{2011A&A...525A..39Y}
\bibinfo{author}{{Yu}, S.} \emph{et~al.}
\newblock \bibinfo{journal}{\bibinfo{title}{{Modelling the radio pulses of an
  ultracool dwarf}}}.
\newblock {\emph{\JournalTitle{\aap}}} \textbf{\bibinfo{volume}{525}},
  \bibinfo{pages}{A39} (\bibinfo{year}{2011}).

\bibitem{2012ApJ...746...99K}
\bibinfo{author}{{Kuznetsov}, A.~A.} \emph{et~al.}
\newblock \bibinfo{journal}{\bibinfo{title}{{Comparative Analysis of Two
  Formation Scenarios of Bursty Radio Emission from Ultracool Dwarfs}}}.
\newblock {\emph{\JournalTitle{\apj}}} \textbf{\bibinfo{volume}{746}},
  \bibinfo{pages}{99} (\bibinfo{year}{2012}).

\bibitem{2019NatCo..10.2276C}
\bibinfo{author}{{Carley}, E.~P.} \emph{et~al.}
\newblock \bibinfo{journal}{\bibinfo{title}{{Loss-cone instability modulation
  due to a magnetohydrodynamic sausage mode oscillation in the solar corona}}}.
\newblock {\emph{\JournalTitle{Nature Communications}}}
  \textbf{\bibinfo{volume}{10}}, \bibinfo{pages}{2276} (\bibinfo{year}{2019}).

\bibitem{2012SoPh..275...17L}
\bibinfo{author}{{Lemen}, J.~R.} \emph{et~al.}
\newblock \bibinfo{journal}{\bibinfo{title}{{The Atmospheric Imaging Assembly
  (AIA) on the Solar Dynamics Observatory (SDO)}}}.
\newblock {\emph{\JournalTitle{\solphys}}} \textbf{\bibinfo{volume}{275}},
  \bibinfo{pages}{17--40} (\bibinfo{year}{2012}).

\bibitem{2020NatAs...4.1140C}
\bibinfo{author}{{Chen}, B.} \emph{et~al.}
\newblock \bibinfo{journal}{\bibinfo{title}{{Measurement of magnetic field and
  relativistic electrons along a solar flare current sheet}}}.
\newblock {\emph{\JournalTitle{Nature Astronomy}}}
  \textbf{\bibinfo{volume}{4}}, \bibinfo{pages}{1140--1147}
  (\bibinfo{year}{2020}).

\bibitem{1994A&A...285..621B}
\bibinfo{author}{{Benz}, A.~O.} \& \bibinfo{author}{{Guedel}, M.}
\newblock \bibinfo{journal}{\bibinfo{title}{{X-ray/microwave ratio of flares
  and coronae}}}.
\newblock {\emph{\JournalTitle{\aap}}} \textbf{\bibinfo{volume}{285}},
  \bibinfo{pages}{621--630} (\bibinfo{year}{1994}).

\bibitem{2006A&ARv..13..229T}
\bibinfo{author}{{Treumann}, R.~A.}
\newblock \bibinfo{journal}{\bibinfo{title}{{The electron-cyclotron maser for
  astrophysical application}}}.
\newblock {\emph{\JournalTitle{\aapr}}} \textbf{\bibinfo{volume}{13}},
  \bibinfo{pages}{229--315} (\bibinfo{year}{2006}).

\bibitem{2018ApJ...866...62C}
\bibinfo{author}{{Chen}, B.} \emph{et~al.}
\newblock \bibinfo{journal}{\bibinfo{title}{{Magnetic Reconnection Null Points
  as the Origin of Semirelativistic Electron Beams in a Solar Jet}}}.
\newblock {\emph{\JournalTitle{\apj}}} \textbf{\bibinfo{volume}{866}},
  \bibinfo{pages}{62} (\bibinfo{year}{2018}).

\bibitem{2021ApJ...922..134B}
\bibinfo{author}{{Battaglia}, M.} \emph{et~al.}
\newblock \bibinfo{journal}{\bibinfo{title}{{Multiple Electron Acceleration
  Instances during a Series of Solar Microflares Observed Simultaneously at
  X-Rays and Microwaves}}}.
\newblock {\emph{\JournalTitle{\apj}}} \textbf{\bibinfo{volume}{922}},
  \bibinfo{pages}{134} (\bibinfo{year}{2021}).

\bibitem{1982Natur.297..485T}
\bibinfo{author}{{Thomas}, J.~H.}, \bibinfo{author}{{Cram}, L.~E.} \&
  \bibinfo{author}{{Nye}, A.~H.}
\newblock \bibinfo{journal}{\bibinfo{title}{{Five-minute oscillations as a
  subsurface probe of sunspot structure}}}.
\newblock {\emph{\JournalTitle{\nat}}} \textbf{\bibinfo{volume}{297}},
  \bibinfo{pages}{485--487} (\bibinfo{year}{1982}).

\bibitem{2023NatAs.tmp..110Y}
\bibinfo{author}{{Yuan}, D.} \emph{et~al.}
\newblock \bibinfo{journal}{\bibinfo{title}{{Transverse oscillations and an
  energy source in a strongly magnetized sunspot}}}.
\newblock {\emph{\JournalTitle{Nature Astronomy}}}  (\bibinfo{year}{2023}).

\bibitem{2018ApJ...864L..24L}
\bibinfo{author}{{Liu}, W.} \emph{et~al.}
\newblock \bibinfo{journal}{\bibinfo{title}{{A Truly Global Extreme Ultraviolet
  Wave from the SOL2017-09-10 X8.2+ Solar Flare-Coronal Mass Ejection}}}.
\newblock {\emph{\JournalTitle{\apjl}}} \textbf{\bibinfo{volume}{864}},
  \bibinfo{pages}{L24} (\bibinfo{year}{2018}).

\bibitem{2017LRSP...14....2B}
\bibinfo{author}{{Benz}, A.~O.}
\newblock \bibinfo{journal}{\bibinfo{title}{{Flare Observations}}}.
\newblock {\emph{\JournalTitle{Living Reviews in Solar Physics}}}
  \textbf{\bibinfo{volume}{14}}, \bibinfo{pages}{2} (\bibinfo{year}{2017}).

\bibitem{2007ApJ...663L.109K}
\bibinfo{author}{{Krucker}, S.}, \bibinfo{author}{{Kontar}, E.~P.},
  \bibinfo{author}{{Christe}, S.} \& \bibinfo{author}{{Lin}, R.~P.}
\newblock \bibinfo{journal}{\bibinfo{title}{{Solar Flare Electron Spectra at
  the Sun and near the Earth}}}.
\newblock {\emph{\JournalTitle{\apjl}}} \textbf{\bibinfo{volume}{663}},
  \bibinfo{pages}{L109--L112} (\bibinfo{year}{2007}).

\bibitem{2019ApJ...871..214V}
\bibinfo{author}{{Villadsen}, J.} \& \bibinfo{author}{{Hallinan}, G.}
\newblock \bibinfo{journal}{\bibinfo{title}{{Ultra-wideband Detection of 22
  Coherent Radio Bursts on M Dwarfs}}}.
\newblock {\emph{\JournalTitle{\apj}}} \textbf{\bibinfo{volume}{871}},
  \bibinfo{pages}{214} (\bibinfo{year}{2019}).

\bibitem{2020NatAs...4..577V}
\bibinfo{author}{{Vedantham}, H.~K.} \emph{et~al.}
\newblock \bibinfo{journal}{\bibinfo{title}{{Coherent radio emission from a
  quiescent red dwarf indicative of star-planet interaction}}}.
\newblock {\emph{\JournalTitle{Nature Astronomy}}}
  \textbf{\bibinfo{volume}{4}}, \bibinfo{pages}{577--583}
  (\bibinfo{year}{2020}).

\bibitem{2021NatAs...5.1233C}
\bibinfo{author}{{Callingham}, J.~R.} \emph{et~al.}
\newblock \bibinfo{journal}{\bibinfo{title}{{The population of M dwarfs
  observed at low radio frequencies}}}.
\newblock {\emph{\JournalTitle{Nature Astronomy}}}
  \textbf{\bibinfo{volume}{5}}, \bibinfo{pages}{1233--1239}
  (\bibinfo{year}{2021}).

\bibitem{2017MNRAS.470.4274T}
\bibinfo{author}{{Turnpenney}, S.}, \bibinfo{author}{{Nichols}, J.~D.},
  \bibinfo{author}{{Wynn}, G.~A.} \& \bibinfo{author}{{Casewell}, S.~L.}
\newblock \bibinfo{journal}{\bibinfo{title}{{Auroral radio emission from
  ultracool dwarfs: a Jovian model}}}.
\newblock {\emph{\JournalTitle{\mnras}}} \textbf{\bibinfo{volume}{470}},
  \bibinfo{pages}{4274--4284} (\bibinfo{year}{2017}).

\bibitem{2022MNRAS.513.1449O}
\bibinfo{author}{{Owocki}, S.~P.} \emph{et~al.}
\newblock \bibinfo{journal}{\bibinfo{title}{{Centrifugal breakout reconnection
  as the electron acceleration mechanism powering the radio magnetospheres of
  early-type stars}}}.
\newblock {\emph{\JournalTitle{\mnras}}} \textbf{\bibinfo{volume}{513}},
  \bibinfo{pages}{1449--1458} (\bibinfo{year}{2022}).

\bibitem{2008AJ....135..785W}
\bibinfo{author}{{West}, A.~A.} \emph{et~al.}
\newblock \bibinfo{journal}{\bibinfo{title}{{Constraining the Age-Activity
  Relation for Cool Stars: The Sloan Digital Sky Survey Data Release 5 Low-Mass
  Star Spectroscopic Sample}}}.
\newblock {\emph{\JournalTitle{\aj}}} \textbf{\bibinfo{volume}{135}},
  \bibinfo{pages}{785--795} (\bibinfo{year}{2008}).

\bibitem{2016ApJ...820...89F}
\bibinfo{author}{{France}, K.} \emph{et~al.}
\newblock \bibinfo{journal}{\bibinfo{title}{{The MUSCLES Treasury Survey. I.
  Motivation and Overview}}}.
\newblock {\emph{\JournalTitle{\apj}}} \textbf{\bibinfo{volume}{820}},
  \bibinfo{pages}{89} (\bibinfo{year}{2016}).

\bibitem{2020ApJ...897..125R}
\bibinfo{author}{{Robertson}, P.} \emph{et~al.}
\newblock \bibinfo{journal}{\bibinfo{title}{{Persistent Starspot Signals on M
  Dwarfs: Multiwavelength Doppler Observations with the Habitable-zone Planet
  Finder and Keck/HIRES}}}.
\newblock {\emph{\JournalTitle{\apj}}} \textbf{\bibinfo{volume}{897}},
  \bibinfo{pages}{125} (\bibinfo{year}{2020}).

\bibitem{2015ApJ...815...64W}
\bibinfo{author}{{Williams}, P.~K.~G.} \emph{et~al.}
\newblock \bibinfo{journal}{\bibinfo{title}{{The First Millimeter Detection of
  a Non-Accreting Ultracool Dwarf}}}.
\newblock {\emph{\JournalTitle{\apj}}} \textbf{\bibinfo{volume}{815}},
  \bibinfo{pages}{64} (\bibinfo{year}{2015}).

\bibitem{2016ApJ...818...24K}
\bibinfo{author}{{Kao}, M.~M.} \emph{et~al.}
\newblock \bibinfo{journal}{\bibinfo{title}{{Auroral Radio Emission from Late L
  and T Dwarfs: A New Constraint on Dynamo Theory in the Substellar Regime}}}.
\newblock {\emph{\JournalTitle{\apj}}} \textbf{\bibinfo{volume}{818}},
  \bibinfo{pages}{24} (\bibinfo{year}{2016}).

\bibitem{2016MNRAS.457.1224L}
\bibinfo{author}{{Lynch}, C.} \emph{et~al.}
\newblock \bibinfo{journal}{\bibinfo{title}{{Radio detections of southern
  ultracool dwarfs}}}.
\newblock {\emph{\JournalTitle{\mnras}}} \textbf{\bibinfo{volume}{457}},
  \bibinfo{pages}{1224--1232} (\bibinfo{year}{2016}).

\bibitem{2017ApJ...846...75P}
\bibinfo{author}{{Pineda}, J.~S.}, \bibinfo{author}{{Hallinan}, G.} \&
  \bibinfo{author}{{Kao}, M.~M.}
\newblock \bibinfo{journal}{\bibinfo{title}{{A Panchromatic View of Brown Dwarf
  Aurorae}}}.
\newblock {\emph{\JournalTitle{\apj}}} \textbf{\bibinfo{volume}{846}},
  \bibinfo{pages}{75} (\bibinfo{year}{2017}).

\bibitem{2018ApJS..237...25K}
\bibinfo{author}{{Kao}, M.~M.}, \bibinfo{author}{{Hallinan}, G.},
  \bibinfo{author}{{Pineda}, J.~S.}, \bibinfo{author}{{Stevenson}, D.} \&
  \bibinfo{author}{{Burgasser}, A.}
\newblock \bibinfo{journal}{\bibinfo{title}{{The Strongest Magnetic Fields on
  the Coolest Brown Dwarfs}}}.
\newblock {\emph{\JournalTitle{\apjs}}} \textbf{\bibinfo{volume}{237}},
  \bibinfo{pages}{25} (\bibinfo{year}{2018}).

\bibitem{2008ApJ...684..644H}
\bibinfo{author}{{Hallinan}, G.} \emph{et~al.}
\newblock \bibinfo{journal}{\bibinfo{title}{{Confirmation of the Electron
  Cyclotron Maser Instability as the Dominant Source of Radio Emission from
  Very Low Mass Stars and Brown Dwarfs}}}.
\newblock {\emph{\JournalTitle{\apj}}} \textbf{\bibinfo{volume}{684}},
  \bibinfo{pages}{644--653} (\bibinfo{year}{2008}).

\bibitem{2017ApJ...836L..30L}
\bibinfo{author}{{Lynch}, C.~R.}, \bibinfo{author}{{Lenc}, E.},
  \bibinfo{author}{{Kaplan}, D.~L.}, \bibinfo{author}{{Murphy}, T.} \&
  \bibinfo{author}{{Anderson}, G.~E.}
\newblock \bibinfo{journal}{\bibinfo{title}{{154 MHz Detection of Faint,
  Polarized Flares from UV Ceti}}}.
\newblock {\emph{\JournalTitle{\apjl}}} \textbf{\bibinfo{volume}{836}},
  \bibinfo{pages}{L30} (\bibinfo{year}{2017}).

\bibitem{2018MNRAS.474L..61D}
\bibinfo{author}{{Das}, B.}, \bibinfo{author}{{Chandra}, P.} \&
  \bibinfo{author}{{Wade}, G.~A.}
\newblock \bibinfo{journal}{\bibinfo{title}{{Discovery of electron cyclotron
  MASER emission from the magnetic Bp star HD 133880 with the Giant Metrewave
  Radio Telescope}}}.
\newblock {\emph{\JournalTitle{\mnras}}} \textbf{\bibinfo{volume}{474}},
  \bibinfo{pages}{L61--L65} (\bibinfo{year}{2018}).

\bibitem{2006ApJ...653..690H}
\bibinfo{author}{{Hallinan}, G.} \emph{et~al.}
\newblock \bibinfo{journal}{\bibinfo{title}{{Rotational Modulation of the Radio
  Emission from the M9 Dwarf TVLM 513-46546: Broadband Coherent Emission at the
  Substellar Boundary?}}}
\newblock {\emph{\JournalTitle{\apj}}} \textbf{\bibinfo{volume}{653}},
  \bibinfo{pages}{690--699} (\bibinfo{year}{2006}).

\bibitem{2008ApJ...673.1080B}
\bibinfo{author}{{Berger}, E.} \emph{et~al.}
\newblock \bibinfo{journal}{\bibinfo{title}{{Simultaneous Multiwavelength
  Observations of Magnetic Activity in Ultracool Dwarfs. I. The Complex
  Behavior of the M8.5 Dwarf TVLM 513-46546}}}.
\newblock {\emph{\JournalTitle{\apj}}} \textbf{\bibinfo{volume}{673}},
  \bibinfo{pages}{1080--1087} (\bibinfo{year}{2008}).

\bibitem{2008ApJ...676.1307B}
\bibinfo{author}{{Berger}, E.} \emph{et~al.}
\newblock \bibinfo{journal}{\bibinfo{title}{{Simultaneous Multiwavelength
  Observations of Magnetic Activity in Ultracool Dwarfs. II. Mixed Trends in VB
  10 and LSR 1835+32 and the Possible Role of Rotation}}}.
\newblock {\emph{\JournalTitle{\apj}}} \textbf{\bibinfo{volume}{676}},
  \bibinfo{pages}{1307--1318} (\bibinfo{year}{2008}).

\bibitem{2011ApJ...739L..10T}
\bibinfo{author}{{Trigilio}, C.}, \bibinfo{author}{{Leto}, P.},
  \bibinfo{author}{{Umana}, G.}, \bibinfo{author}{{Buemi}, C.~S.} \&
  \bibinfo{author}{{Leone}, F.}
\newblock \bibinfo{journal}{\bibinfo{title}{{Auroral Radio Emission from Stars:
  The Case of CU Virginis}}}.
\newblock {\emph{\JournalTitle{\apjl}}} \textbf{\bibinfo{volume}{739}},
  \bibinfo{pages}{L10} (\bibinfo{year}{2011}).

\bibitem{2000ApJ...538..456E}
\bibinfo{author}{{Ergun}, R.~E.} \emph{et~al.}
\newblock \bibinfo{journal}{\bibinfo{title}{{Electron-Cyclotron Maser Driven by
  Charged-Particle Acceleration from Magnetic Field-aligned Electric Fields}}}.
\newblock {\emph{\JournalTitle{\apj}}} \textbf{\bibinfo{volume}{538}},
  \bibinfo{pages}{456--466} (\bibinfo{year}{2000}).

\bibitem{2005SoPh..227..231W}
\bibinfo{author}{{White}, S.~M.}, \bibinfo{author}{{Thomas}, R.~J.} \&
  \bibinfo{author}{{Schwartz}, R.~A.}
\newblock \bibinfo{journal}{\bibinfo{title}{{Updated Expressions for
  Determining Temperatures and Emission Measures from Goes Soft X-Ray
  Measurements}}}.
\newblock {\emph{\JournalTitle{\solphys}}} \textbf{\bibinfo{volume}{227}},
  \bibinfo{pages}{231--248} (\bibinfo{year}{2005}).

\bibitem{2013PhDT.......498C}
\bibinfo{author}{{Chen}, B.}
\newblock \emph{\bibinfo{title}{{Radio and X-ray diagnostics of energy release
  in solar flares}}}.
\newblock Ph.D. thesis, \bibinfo{school}{University of Virginia}
  (\bibinfo{year}{2013}).

\bibitem{1997PASP..109..166C}
\bibinfo{author}{{Condon}, J.~J.}
\newblock \bibinfo{journal}{\bibinfo{title}{{Errors in Elliptical Gaussian
  Fits}}}.
\newblock {\emph{\JournalTitle{\pasp}}} \textbf{\bibinfo{volume}{109}},
  \bibinfo{pages}{166--172} (\bibinfo{year}{1997}).

\bibitem{1994ApJ...426..774B}
\bibinfo{author}{{Bastian}, T.~S.}
\newblock \bibinfo{journal}{\bibinfo{title}{{Angular Scattering of Solar Radio
  Emission by Coronal Turbulence}}}.
\newblock {\emph{\JournalTitle{\apj}}} \textbf{\bibinfo{volume}{426}},
  \bibinfo{pages}{774} (\bibinfo{year}{1994}).

\bibitem{2017NatCo...8.1515K}
\bibinfo{author}{{Kontar}, E.~P.} \emph{et~al.}
\newblock \bibinfo{journal}{\bibinfo{title}{{Imaging spectroscopy of solar
  radio burst fine structures}}}.
\newblock {\emph{\JournalTitle{Nature Communications}}}
  \textbf{\bibinfo{volume}{8}}, \bibinfo{pages}{1515} (\bibinfo{year}{2017}).

\bibitem{1985PASJ...37..163N}
\bibinfo{author}{{Nakajima}, H.}, \bibinfo{author}{{Sekiguchi}, H.},
  \bibinfo{author}{{Sawa}, M.}, \bibinfo{author}{{Kai}, K.} \&
  \bibinfo{author}{{Kawashima}, S.}
\newblock \bibinfo{journal}{\bibinfo{title}{{The radiometer and polarimeters at
  80, 35, and 17 GHz for solar observations at Nobeyama}}}.
\newblock {\emph{\JournalTitle{\pasj}}} \textbf{\bibinfo{volume}{37}},
  \bibinfo{pages}{163--170} (\bibinfo{year}{1985}).

\bibitem{2010ApJ...709..332B}
\bibinfo{author}{{Berger}, E.} \emph{et~al.}
\newblock \bibinfo{journal}{\bibinfo{title}{{Simultaneous Multi-Wavelength
  Observations of Magnetic Activity in Ultracool Dwarfs. III. X-ray, Radio, and
  H{\ensuremath{\alpha}} Activity Trends in M and L dwarfs}}}.
\newblock {\emph{\JournalTitle{\apj}}} \textbf{\bibinfo{volume}{709}},
  \bibinfo{pages}{332--341} (\bibinfo{year}{2010}).

\bibitem{2014ApJ...785....9W}
\bibinfo{author}{{Williams}, P.~K.~G.}, \bibinfo{author}{{Cook}, B.~A.} \&
  \bibinfo{author}{{Berger}, E.}
\newblock \bibinfo{journal}{\bibinfo{title}{{Trends in Ultracool Dwarf
  Magnetism. I. X-Ray Suppression and Radio Enhancement}}}.
\newblock {\emph{\JournalTitle{\apj}}} \textbf{\bibinfo{volume}{785}},
  \bibinfo{pages}{9} (\bibinfo{year}{2014}).

\bibitem{1977ApJS...35..419R}
\bibinfo{author}{{Raymond}, J.~C.} \& \bibinfo{author}{{Smith}, B.~W.}
\newblock \bibinfo{journal}{\bibinfo{title}{{Soft X-ray spectrum of a hot
  plasma.}}}
\newblock {\emph{\JournalTitle{\apjs}}} \textbf{\bibinfo{volume}{35}},
  \bibinfo{pages}{419--439} (\bibinfo{year}{1977}).

\bibitem{2004SoPh..219...87W}
\bibinfo{author}{{Wiegelmann}, T.}
\newblock \bibinfo{journal}{\bibinfo{title}{{Optimization code with weighting
  function for the reconstruction of coronal magnetic fields}}}.
\newblock {\emph{\JournalTitle{\solphys}}} \textbf{\bibinfo{volume}{219}},
  \bibinfo{pages}{87--108} (\bibinfo{year}{2004}).

\bibitem{1995ApJ...439..474M}
\bibinfo{author}{{Metcalf}, T.~R.}, \bibinfo{author}{{Jiao}, L.},
  \bibinfo{author}{{McClymont}, A.~N.}, \bibinfo{author}{{Canfield}, R.~C.} \&
  \bibinfo{author}{{Uitenbroek}, H.}
\newblock \bibinfo{journal}{\bibinfo{title}{{Is the Solar Chromospheric
  Magnetic Field Force-free?}}}
\newblock {\emph{\JournalTitle{\apj}}} \textbf{\bibinfo{volume}{439}},
  \bibinfo{pages}{474} (\bibinfo{year}{1995}).

\bibitem{2012SoPh..275..207S}
\bibinfo{author}{{Scherrer}, P.~H.} \emph{et~al.}
\newblock \bibinfo{journal}{\bibinfo{title}{{The Helioseismic and Magnetic
  Imager (HMI) Investigation for the Solar Dynamics Observatory (SDO)}}}.
\newblock {\emph{\JournalTitle{\solphys}}} \textbf{\bibinfo{volume}{275}},
  \bibinfo{pages}{207--227} (\bibinfo{year}{2012}).

\bibitem{2014SoPh..289.3549B}
\bibinfo{author}{{Bobra}, M.~G.} \emph{et~al.}
\newblock \bibinfo{journal}{\bibinfo{title}{{The Helioseismic and Magnetic
  Imager (HMI) Vector Magnetic Field Pipeline: SHARPs - Space-Weather HMI
  Active Region Patches}}}.
\newblock {\emph{\JournalTitle{\solphys}}} \textbf{\bibinfo{volume}{289}},
  \bibinfo{pages}{3549--3578} (\bibinfo{year}{2014}).

\bibitem{2006SoPh..233..215W}
\bibinfo{author}{{Wiegelmann}, T.}, \bibinfo{author}{{Inhester}, B.} \&
  \bibinfo{author}{{Sakurai}, T.}
\newblock \bibinfo{journal}{\bibinfo{title}{{Preprocessing of Vector
  Magnetograph Data for a Nonlinear Force-Free Magnetic Field
  Reconstruction}}}.
\newblock {\emph{\JournalTitle{\solphys}}} \textbf{\bibinfo{volume}{233}},
  \bibinfo{pages}{215--232} (\bibinfo{year}{2006}).

\bibitem{2006A&A...449..791T}
\bibinfo{author}{{Thompson}, W.~T.}
\newblock \bibinfo{journal}{\bibinfo{title}{{Coordinate systems for solar image
  data}}}.
\newblock {\emph{\JournalTitle{\aap}}} \textbf{\bibinfo{volume}{449}},
  \bibinfo{pages}{791--803} (\bibinfo{year}{2006}).

\bibitem{2000asqu.book.....C}
\bibinfo{author}{{Cox}, A.~N.}
\newblock \emph{\bibinfo{title}{{Allen's astrophysical quantities}}}
  (\bibinfo{year}{2000}).

\bibitem{2012A&A...539A.146H}
\bibinfo{author}{{Hannah}, I.~G.} \& \bibinfo{author}{{Kontar}, E.~P.}
\newblock \bibinfo{journal}{\bibinfo{title}{{Differential emission measures
  from the regularized inversion of Hinode and SDO data}}}.
\newblock {\emph{\JournalTitle{\aap}}} \textbf{\bibinfo{volume}{539}},
  \bibinfo{pages}{A146} (\bibinfo{year}{2012}).

\bibitem{1982ApJ...259..844M}
\bibinfo{author}{{Melrose}, D.~B.} \& \bibinfo{author}{{Dulk}, G.~A.}
\newblock \bibinfo{journal}{\bibinfo{title}{{Electron-cyclotron masers as the
  source of certain solar and stellar radio bursts.}}}
\newblock {\emph{\JournalTitle{\apj}}} \textbf{\bibinfo{volume}{259}},
  \bibinfo{pages}{844--858} (\bibinfo{year}{1982}).

\bibitem{1979ApJ...230..621W}
\bibinfo{author}{{Wu}, C.~S.} \& \bibinfo{author}{{Lee}, L.~C.}
\newblock \bibinfo{journal}{\bibinfo{title}{{A theory of the terrestrial
  kilometric radiation.}}}
\newblock {\emph{\JournalTitle{\apj}}} \textbf{\bibinfo{volume}{230}},
  \bibinfo{pages}{621--626} (\bibinfo{year}{1979}).

\bibitem{2007P&SS...55...89H}
\bibinfo{author}{{Hess}, S.}, \bibinfo{author}{{Zarka}, P.} \&
  \bibinfo{author}{{Mottez}, F.}
\newblock \bibinfo{journal}{\bibinfo{title}{{Io Jupiter interaction,
  millisecond bursts and field-aligned potentials}}}.
\newblock {\emph{\JournalTitle{\planss}}} \textbf{\bibinfo{volume}{55}},
  \bibinfo{pages}{89--99} (\bibinfo{year}{2007}).

\bibitem{2008JGRA..113.7201L}
\bibinfo{author}{{Lamy}, L.} \emph{et~al.}
\newblock \bibinfo{journal}{\bibinfo{title}{{Saturn kilometric radiation:
  Average and statistical properties}}}.
\newblock {\emph{\JournalTitle{Journal of Geophysical Research (Space
  Physics)}}} \textbf{\bibinfo{volume}{113}}, \bibinfo{pages}{A07201}
  (\bibinfo{year}{2008}).

\bibitem{2015A&A...581A...9R}
\bibinfo{author}{{R{\'e}gnier}, S.}
\newblock \bibinfo{journal}{\bibinfo{title}{{A new approach to the maser
  emission in the solar corona}}}.
\newblock {\emph{\JournalTitle{\aap}}} \textbf{\bibinfo{volume}{581}},
  \bibinfo{pages}{A9} (\bibinfo{year}{2015}).

\bibitem{2016A&A...589L...8M}
\bibinfo{author}{{Morosan}, D.~E.}, \bibinfo{author}{{Zucca}, P.},
  \bibinfo{author}{{Bloomfield}, D.~S.} \& \bibinfo{author}{{Gallagher}, P.~T.}
\newblock \bibinfo{journal}{\bibinfo{title}{{Conditions for electron-cyclotron
  maser emission in the solar corona}}}.
\newblock {\emph{\JournalTitle{\aap}}} \textbf{\bibinfo{volume}{589}},
  \bibinfo{pages}{L8} (\bibinfo{year}{2016}).

\bibitem{1968SvA....12..245Z}
\bibinfo{author}{{Zlotnik}, E.~Y.}
\newblock \bibinfo{journal}{\bibinfo{title}{{Theory of the Slowly Changing
  Component of Solar Radio Emission. I.}}}
\newblock {\emph{\JournalTitle{\sovast}}} \textbf{\bibinfo{volume}{12}},
  \bibinfo{pages}{245} (\bibinfo{year}{1968}).

\bibitem{1985JGR....90.9663W}
\bibinfo{author}{{Winglee}, R.~M.}
\newblock \bibinfo{journal}{\bibinfo{title}{{Fundamental and harmonic electron
  cyclotron maser emission}}}.
\newblock {\emph{\JournalTitle{\jgr}}} \textbf{\bibinfo{volume}{90}},
  \bibinfo{pages}{9663--9674} (\bibinfo{year}{1985}).

\bibitem{2017PhPl...24e2902T}
\bibinfo{author}{{Tong}, Z.-J.}, \bibinfo{author}{{Wang}, C.-B.},
  \bibinfo{author}{{Zhang}, P.-J.} \& \bibinfo{author}{{Liu}, J.}
\newblock \bibinfo{journal}{\bibinfo{title}{{A parametric investigation on the
  cyclotron maser instability driven by ring-beam electrons with intrinsic
  Alfv{\'e}n waves}}}.
\newblock {\emph{\JournalTitle{Physics of Plasmas}}}
  \textbf{\bibinfo{volume}{24}}, \bibinfo{pages}{052902}
  (\bibinfo{year}{2017}).

\bibitem{2013ApJ...763L..21C}
\bibinfo{author}{{Chen}, B.} \emph{et~al.}
\newblock \bibinfo{journal}{\bibinfo{title}{{Tracing Electron Beams in the
  Sun's Corona with Radio Dynamic Imaging Spectroscopy}}}.
\newblock {\emph{\JournalTitle{\apjl}}} \textbf{\bibinfo{volume}{763}},
  \bibinfo{pages}{L21} (\bibinfo{year}{2013}).

\bibitem{2008JGRA..113.3209H}
\bibinfo{author}{{Hess}, S.}, \bibinfo{author}{{Mottez}, F.},
  \bibinfo{author}{{Zarka}, P.} \& \bibinfo{author}{{Chust}, T.}
\newblock \bibinfo{journal}{\bibinfo{title}{{Generation of the jovian radio
  decametric arcs from the Io Flux Tube}}}.
\newblock {\emph{\JournalTitle{Journal of Geophysical Research: Space
  Physics}}} \textbf{\bibinfo{volume}{113}}, \bibinfo{pages}{A03209}
  (\bibinfo{year}{2008}).

\bibitem{2012PhPl...19h2902W}
\bibinfo{author}{{Wu}, C.~S.}, \bibinfo{author}{{Wang}, C.~B.},
  \bibinfo{author}{{Wu}, D.~J.} \& \bibinfo{author}{{Lee}, K.~H.}
\newblock \bibinfo{journal}{\bibinfo{title}{{Resonant wave-particle
  interactions modified by intrinsic Alfv{\'e}nic turbulence}}}.
\newblock {\emph{\JournalTitle{Physics of Plasmas}}}
  \textbf{\bibinfo{volume}{19}}, \bibinfo{pages}{082902}
  (\bibinfo{year}{2012}).

\bibitem{2011Natur.475..477M}
\bibinfo{author}{{McIntosh}, S.~W.} \emph{et~al.}
\newblock \bibinfo{journal}{\bibinfo{title}{{Alfv{\'e}nic waves with sufficient
  energy to power the quiet solar corona and fast solar wind}}}.
\newblock {\emph{\JournalTitle{\nat}}} \textbf{\bibinfo{volume}{475}},
  \bibinfo{pages}{477--480} (\bibinfo{year}{2011}).

\bibitem{2018NatPh..14..480G}
\bibinfo{author}{{Grant}, S. D.~T.} \emph{et~al.}
\newblock \bibinfo{journal}{\bibinfo{title}{{Alfv{\'e}n wave dissipation in the
  solar chromosphere}}}.
\newblock {\emph{\JournalTitle{Nature Physics}}} \textbf{\bibinfo{volume}{14}},
  \bibinfo{pages}{480--483} (\bibinfo{year}{2018}).

\bibitem{2019ApJ...872...71Y}
\bibinfo{author}{{Yu}, S.} \& \bibinfo{author}{{Chen}, B.}
\newblock \bibinfo{journal}{\bibinfo{title}{{Possible Detection of
  Subsecond-period Propagating Magnetohydrodynamics Waves in Post-reconnection
  Magnetic Loops during a Two-ribbon Solar Flare}}}.
\newblock {\emph{\JournalTitle{\apj}}} \textbf{\bibinfo{volume}{872}},
  \bibinfo{pages}{71} (\bibinfo{year}{2019}).

\bibitem{2005psci.book.....A}
\bibinfo{author}{{Aschwanden}, M.~J.}
\newblock \emph{\bibinfo{title}{{Physics of the Solar Corona. An Introduction
  with Problems and Solutions (2nd edition)}}} (\bibinfo{year}{2005}).

\bibitem{2019NatAs...3..160C}
\bibinfo{author}{{Cheung}, M.~C.~M.} \emph{et~al.}
\newblock \bibinfo{journal}{\bibinfo{title}{{A comprehensive three-dimensional
  radiative magnetohydrodynamic simulation of a solar flare}}}.
\newblock {\emph{\JournalTitle{Nature Astronomy}}}
  \textbf{\bibinfo{volume}{3}}, \bibinfo{pages}{160--166}
  (\bibinfo{year}{2019}).

\bibitem{2007ApJ...668.1210R}
\bibinfo{author}{{Reeves}, K.~K.}, \bibinfo{author}{{Warren}, H.~P.} \&
  \bibinfo{author}{{Forbes}, T.~G.}
\newblock \bibinfo{journal}{\bibinfo{title}{{Theoretical Predictions of X-Ray
  and Extreme-UV Flare Emissions Using a Loss-of-Equilibrium Model of Solar
  Eruptions}}}.
\newblock {\emph{\JournalTitle{\apj}}} \textbf{\bibinfo{volume}{668}},
  \bibinfo{pages}{1210--1220} (\bibinfo{year}{2007}).

\bibitem{2015SoPh..290.3457S}
\bibinfo{author}{{Schmieder}, B.}, \bibinfo{author}{{Aulanier}, G.} \&
  \bibinfo{author}{{Vr{\v{s}}nak}, B.}
\newblock \bibinfo{journal}{\bibinfo{title}{{Flare-CME Models: An Observational
  Perspective (Invited Review)}}}.
\newblock {\emph{\JournalTitle{\solphys}}} \textbf{\bibinfo{volume}{290}},
  \bibinfo{pages}{3457--3486} (\bibinfo{year}{2015}).

\bibitem{2022ApJ...925..125D}
\bibinfo{author}{{Das}, B.} \emph{et~al.}
\newblock \bibinfo{journal}{\bibinfo{title}{{Discovery of Eight ``Main-sequence
  Radio Pulse Emitters'' Using the GMRT: Clues to the Onset of Coherent Radio
  Emission in Hot Magnetic Stars}}}.
\newblock {\emph{\JournalTitle{\apj}}} \textbf{\bibinfo{volume}{925}},
  \bibinfo{pages}{125} (\bibinfo{year}{2022}).

\bibitem{2004A&A...418..593T}
\bibinfo{author}{{Trigilio}, C.}, \bibinfo{author}{{Leto}, P.},
  \bibinfo{author}{{Umana}, G.}, \bibinfo{author}{{Leone}, F.} \&
  \bibinfo{author}{{Buemi}, C.~S.}
\newblock \bibinfo{journal}{\bibinfo{title}{{A three-dimensional model for the
  radio emission of magnetic chemically peculiar stars}}}.
\newblock {\emph{\JournalTitle{\aap}}} \textbf{\bibinfo{volume}{418}},
  \bibinfo{pages}{593--605} (\bibinfo{year}{2004}).

\bibitem{2019A&A...621A..47K}
\bibinfo{author}{{Kochukhov}, O.}, \bibinfo{author}{{Shultz}, M.} \&
  \bibinfo{author}{{Neiner}, C.}
\newblock \bibinfo{journal}{\bibinfo{title}{{Magnetic field topologies of the
  bright, weak-field Ap stars {\ensuremath{\theta}} Aurigae and ? Ursae
  Majoris}}}.
\newblock {\emph{\JournalTitle{\aap}}} \textbf{\bibinfo{volume}{621}},
  \bibinfo{pages}{A47} (\bibinfo{year}{2019}).

\bibitem{2020MNRAS.493.4657L}
\bibinfo{author}{{Leto}, P.} \emph{et~al.}
\newblock \bibinfo{journal}{\bibinfo{title}{{Evidence for radio and X-ray
  auroral emissions from the magnetic B-type star {\ensuremath{\rho}} Oph A}}}.
\newblock {\emph{\JournalTitle{\mnras}}} \textbf{\bibinfo{volume}{493}},
  \bibinfo{pages}{4657--4676} (\bibinfo{year}{2020}).

\bibitem{2014A&A...565A..83K}
\bibinfo{author}{{Kochukhov}, O.}, \bibinfo{author}{{L{\"u}ftinger}, T.},
  \bibinfo{author}{{Neiner}, C.}, \bibinfo{author}{{Alecian}, E.} \&
  \bibinfo{author}{{MiMeS Collaboration}}.
\newblock \bibinfo{journal}{\bibinfo{title}{{Magnetic field topology of the
  unique chemically peculiar star CU Virginis}}}.
\newblock {\emph{\JournalTitle{\aap}}} \textbf{\bibinfo{volume}{565}},
  \bibinfo{pages}{A83} (\bibinfo{year}{2014}).

\bibitem{2019MNRAS.485.3457B}
\bibinfo{author}{{Balona}, L.~A.} \emph{et~al.}
\newblock \bibinfo{journal}{\bibinfo{title}{{Rotational modulation in TESS B
  stars}}}.
\newblock {\emph{\JournalTitle{\mnras}}} \textbf{\bibinfo{volume}{485}},
  \bibinfo{pages}{3457--3469} (\bibinfo{year}{2019}).

\bibitem{2011A&A...534A.140C}
\bibinfo{author}{{Cantiello}, M.} \& \bibinfo{author}{{Braithwaite}, J.}
\newblock \bibinfo{journal}{\bibinfo{title}{{Magnetic spots on hot massive
  stars}}}.
\newblock {\emph{\JournalTitle{\aap}}} \textbf{\bibinfo{volume}{534}},
  \bibinfo{pages}{A140} (\bibinfo{year}{2011}).

\bibitem{2004A&A...422..225M}
\bibinfo{author}{{Maeder}, A.} \& \bibinfo{author}{{Meynet}, G.}
\newblock \bibinfo{journal}{\bibinfo{title}{{Stellar evolution with rotation
  and magnetic fields. II. General equations for the transport by Tayler-Spruit
  dynamo}}}.
\newblock {\emph{\JournalTitle{\aap}}} \textbf{\bibinfo{volume}{422}},
  \bibinfo{pages}{225--237} (\bibinfo{year}{2004}).

\bibitem{2018MNRAS.475.5144S}
\bibinfo{author}{{Shultz}, M.~E.} \emph{et~al.}
\newblock \bibinfo{journal}{\bibinfo{title}{{The magnetic early B-type stars I:
  magnetometry and rotation}}}.
\newblock {\emph{\JournalTitle{\mnras}}} \textbf{\bibinfo{volume}{475}},
  \bibinfo{pages}{5144--5178} (\bibinfo{year}{2018}).

\bibitem{1986SoPh..104...99B}
\bibinfo{author}{{Benz}, A.~O.}
\newblock \bibinfo{journal}{\bibinfo{title}{{Millisecond Radio Spikes}}}.
\newblock {\emph{\JournalTitle{\solphys}}} \textbf{\bibinfo{volume}{104}},
  \bibinfo{pages}{99--110} (\bibinfo{year}{1986}).

\bibitem{1996ApJ...459L..95J}
\bibinfo{author}{{Johns-Krull}, C.~M.} \& \bibinfo{author}{{Valenti}, J.~A.}
\newblock \bibinfo{journal}{\bibinfo{title}{{Detection of Strong Magnetic
  Fields on M Dwarfs}}}.
\newblock {\emph{\JournalTitle{\apjl}}} \textbf{\bibinfo{volume}{459}},
  \bibinfo{pages}{L95} (\bibinfo{year}{1996}).

\bibitem{2010MNRAS.407.2269M}
\bibinfo{author}{{Morin}, J.} \emph{et~al.}
\newblock \bibinfo{journal}{\bibinfo{title}{{Large-scale magnetic topologies of
  late M dwarfs*}}}.
\newblock {\emph{\JournalTitle{\mnras}}} \textbf{\bibinfo{volume}{407}},
  \bibinfo{pages}{2269--2286} (\bibinfo{year}{2010}).

\bibitem{2021A&ARv..29....1K}
\bibinfo{author}{{Kochukhov}, O.}
\newblock \bibinfo{journal}{\bibinfo{title}{{Magnetic fields of M dwarfs}}}.
\newblock {\emph{\JournalTitle{\aapr}}} \textbf{\bibinfo{volume}{29}},
  \bibinfo{pages}{1} (\bibinfo{year}{2021}).

\bibitem{2007ASPC..376..127M}
\bibinfo{author}{{McMullin}, J.~P.}, \bibinfo{author}{{Waters}, B.},
  \bibinfo{author}{{Schiebel}, D.}, \bibinfo{author}{{Young}, W.} \&
  \bibinfo{author}{{Golap}, K.}
\newblock \bibinfo{title}{{CASA Architecture and Applications}}.
\newblock In \bibinfo{editor}{{Shaw}, R.~A.}, \bibinfo{editor}{{Hill}, F.} \&
  \bibinfo{editor}{{Bell}, D.~J.} (eds.) \emph{\bibinfo{booktitle}{Astronomical
  Data Analysis Software and Systems XVI}}, vol. \bibinfo{volume}{376} of
  \emph{\bibinfo{series}{Astronomical Society of the Pacific Conference
  Series}}, \bibinfo{pages}{127} (\bibinfo{year}{2007}).

\bibitem{2020ApJ...890...68S}
\bibinfo{author}{{SunPy Community}} \emph{et~al.}
\newblock \bibinfo{journal}{\bibinfo{title}{{The SunPy Project: Open Source
  Development and Status of the Version 1.0 Core Package}}}.
\newblock {\emph{\JournalTitle{\apj}}} \textbf{\bibinfo{volume}{890}},
  \bibinfo{pages}{68} (\bibinfo{year}{2020}).

\bibitem{2022ApJ...935..167A}
\bibinfo{author}{{Astropy Collaboration}} \emph{et~al.}
\newblock \bibinfo{journal}{\bibinfo{title}{{The Astropy Project: Sustaining
  and Growing a Community-oriented Open-source Project and the Latest Major
  Release (v5.0) of the Core Package}}}.
\newblock {\emph{\JournalTitle{\apj}}} \textbf{\bibinfo{volume}{935}},
  \bibinfo{pages}{167} (\bibinfo{year}{2022}).

\end{thebibliography}

\section*{Acknowledgements}
The VLA is operated by the National Radio Astronomy Observatory (NRAO), a facility of the National Science Foundation (NSF) operated under a cooperative agreement by Associated Universities, Inc. S.Y. and B.C. are supported by 
NASA HSOC and ECIP grants
(80NSSC20K1283/SV0-09025 %HSO-connect
and 80NSSC21K0623), % ECIP
% and 80NSSC20K1318, % HSR SOSS
and NSF grant AGS-1654382 
to NJIT. 
R.S. acknowledges the support of the Swiss National Foundation, under grant 200021\_175832. 
R.S. also acknowledges Dr. S\"{a}m Krucker and Dr. Andr{\'e} Csillaghy, FHNW, for their support.
D.E.G. acknowledges support from NASA HSR grant 80NSSC18K1128.
The authors acknowledge Dr. Arnold Benz for the useful discussions.

\section*{Author contributions statement}
S.Y. conceived the study, carried out the data reduction, analysis, visualization, interpretation and manuscript preparation. B.C. led the VLA observing proposal (VLA/16A-377) and planned and conducted the VLA observations. T.S.B. and D.E.G. participated in the VLA proposal and the observing programs. 
R.S. conducted an independent VLA data reduction and cross checked the results. 
B.C., R.S., T.S.B., S.M., and D.E.G. contributed to the theoretical interpretation.
S.Y. and B.C. led the manuscript writing, and all authors discussed the results and commented on the manuscript.

\section*{Competing Interests Statement}
The authors declare no competing interests.

\end{document}